\documentclass{natureprintstyle}

\usepackage{psfig}

\usepackage{lscape}
\usepackage{amsmath}
\usepackage{amssymb}
\usepackage{color}
\usepackage{url}
\usepackage{ulem}
\usepackage{multirow}
\usepackage{graphics,graphicx}
\usepackage[affil-it]{authblk}
\usepackage{blindtext}
\usepackage{subfig}
\usepackage{hyperref}

\usepackage{astjnlabbrev-nature} 

\definecolor{color1}{rgb}{0.0,0.0,.0} % Color of the title sections
\definecolor{color2}{rgb}{0.925, 0.956, 0.992} % Color of the boxes behind the headings
\definecolor{color3}{rgb}{1.0,1.0,1.0} % Color of the boxes behind the abstract 
\definecolor{color4}{rgb}{0.0,0.0,0.0} % Color of the title
\definecolor{colorbox}{rgb}{0.925, 0.956, 0.992}
\definecolor{colorheader}{rgb}{0.33,0.41,0.47} 
\usepackage{hyperref} % Required for hyperlinks
\hypersetup{hidelinks,colorlinks,breaklinks=true,urlcolor=color2,citecolor=color1,linkcolor=color1,bookmarksopen=false,pdftitle={Title},pdfauthor={Author}}
\def\degr{^{\rm o}}

\def\pasa{{\it Publ. Astron. Soc. Aust.}}
\def\nar{{\it New Astron. Rev.}}
\def\mps{{\it Meteorit. Planet. Sci.}}

\def\spose#1{\hbox to 0pt{#1\hss}}
\newcommand       \Angstrom     {\,{\rm \AA}}

\newcommand       \K            {\,{\rm K}}
\newcommand       \pc           {\,{\rm pc}}

\newcommand       \yr       {\,{\rm yr}}

\newcommand       \Myr      {\,{\rm Myr}}

\newcommand       \Gyr      {\,{\rm Gyr}}

\newcommand       \Msun   {M_{\odot}}

\newcommand     \gtsim  {\lower.5ex\hbox{$\buildrel > \over \sim$}}
\newcommand     \ltsim  {\lower.5ex\hbox{$\buildrel < \over \sim$}}
\newcommand     \simgt  {\lower.5ex\hbox{$\buildrel > \over \sim$}}
\newcommand     \simlt  {\lower.5ex\hbox{$\buildrel < \over \sim$}}

\newcommand       \mum          {\,{\rm \mu m}}

\newcommand       \Teff         {T_{\star}}

\newcommand       \simali       {\sim\,}

\makeatletter
\renewcommand{\section}{\@startsection%
{section}{1}{0mm}{-\baselineskip}%
{0.5\baselineskip}{\normalfont\Large\bfseries}}%
\makeatother

\title{{\it Spitzer}'s perspective of polycyclic aromatic hydrocarbons
  in galaxies}
\author{Aigen~Li}
\begin{document}
\pagestyle{plain}
\pagenumbering{arabic}
%\noindent

\twocolumn[
  \begin{@twocolumnfalse}

\let\newpage\relax\maketitle

\begin{affiliations}
%\item
Department of Physics and Astronomy,
University of Missouri,
Columbia, MO 65211, USA.
e-mail: {\sf lia@missouri.edu}
\end{affiliations}

\bigskip

\begin{abstract}
Polycyclic aromatic hydrocarbon (PAH) molecules,
as revealed by the distinctive set of emission bands 
at 3.3, 6.2, 7.7, 8.6, 11.3 and 12.7$\mum$ 
characteristic of their vibrational modes,
are abundant and widespread throughout the Universe.
They are ubiquitously seen in a wide variety of
astrophysical regions, ranging from planet-forming 
disks around young stars to the interstellar medium (ISM) 
of the Milky Way and external galaxies 
out to high redshifts at $z$\,$\simgt$\,4.
PAHs profoundly influence the thermal budget 
and chemistry of the ISM by dominating
the photoelectric heating of the gas
and controlling the ionization balance.
Here, we review the current state of knowledge 
of the astrophysics of PAHs, focusing on their 
observational characteristics obtained from 
the {\it Spitzer Space Telescope} 
and their diagnostic power for probing 
the local physical and chemical conditions
and processes. Special attention is paid to
the spectral properties of PAHs and their variations  
revealed by the {\it Infrared Spectrograph} (IRS)
on board {\it Spitzer} across a much broader range 
of extragalactic environments (e.g., distant galaxies,
early-type galaxies, galactic halos, active galactic nuclei,
and low-metallicity galaxies) than was previously 
possible with the {\it Infrared Space Observatory} 
(ISO) or any other telescope facilities. 
Also highlighted is the relation between 
the PAH abundance and the galaxy metallicity 
established for the first time by {\it Spitzer}.

\bigskip
\end{abstract}
\end{@twocolumnfalse}
]
%{
%  \renewcommand{\thefootnote}%
%    {\fnsymbol{footnote}}
%  \footnotetext[1]{\small %The authors' order is purely 
%  alphabetical since they both have contributed equally 
%  to this Review. e-mail: lia@missouri.edu}
%}

\begin{figure*}[t!]
\centering
\includegraphics[width=14cm,height=10cm]{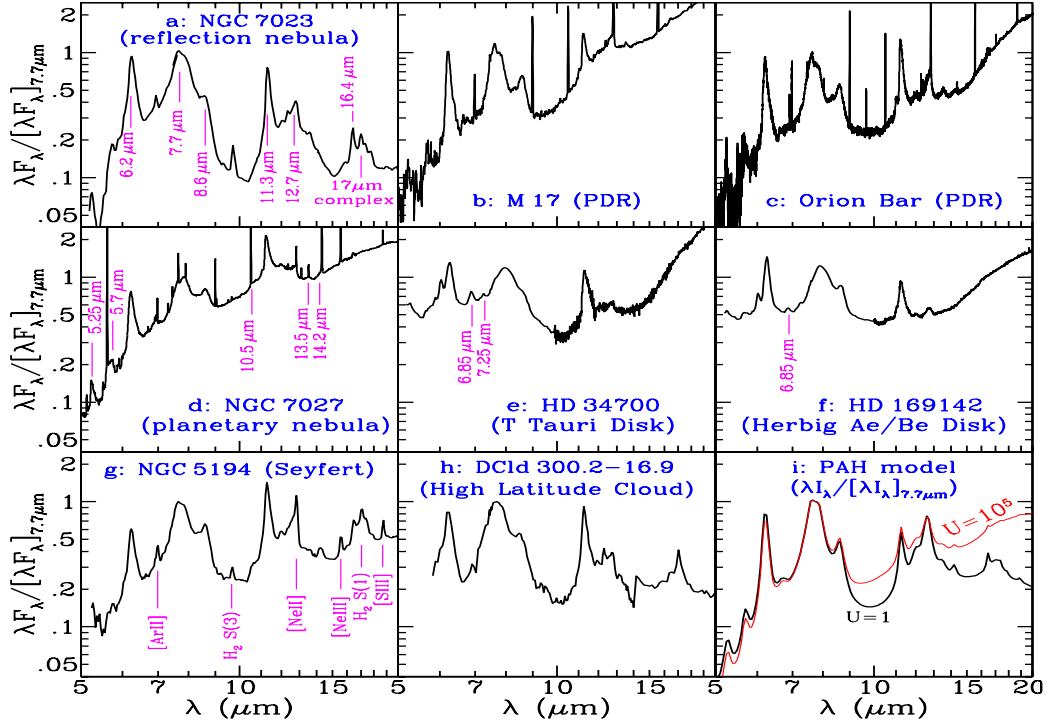}
%\hspace{-2mm}
%\includegraphics[width=8.2cm,height=7.5cm]{pahspec.ps}
\caption{{\bf Observed and model-predicted
               5--20$\mum$ PAH spectra.}
	 (a) Reflection nebula NGC\,7023 (ref.\cite{Werner2004});
	 (b) M17 PDR\cite{Peeters2005};
         (c) Orion Bar PDR\cite{Verstraete2001};
	 (d) Planetary nebula NGC\,7027
         (ref.\cite{vanDiedenhoven2004});
         (e) T Tauri disk HD\,34700 (ref.\cite{Schuetz2009});
         (f) Herbig Ae/Be disk HD\,169142 (ref.\cite{Sloan2005});
	 (g) Seyfert grand-design spiral 
              galaxy NGC\,5194 (ref.\cite{Smith2007});
         (h) Translucent high Galactic latitude cloud
              DCld 300.2-16.9 (ref.\cite{Ingalls2011});
         (i) Model emission calculated for
	 PAHs illuminated by the local interstellar 
         radiation field\cite{MMP83} 
         (i.e., $U$\,=\,1; black line) 
         or a much more intense radiation field 
         (i.e., $U$\,=\,10$^5$; red line)\cite{DL07}.
         The major PAH bands are labelled in (a).
         Some of the weak, secondary PAH bands
         (superimposed by sharp gas lines)
         are labelled in (d). 
         The 6.85 and 7.25$\mum$ aliphatic 
         C--H deformation bands are labelled
         in (e) and (f) for protoplanetary disks.
         The sharp gaseous emission lines 
         are labelled in (g)\cite{Seok2015,Seok2016}.
         %Figure adapted from ref.\cite{DL07}.
         }
\label{fig:pahspec}
\end{figure*}

%%%%%% main text %%%%%%
In the early 1970s,  a new chapter in astrochemistry 
was opened first by Gillett et al.\cite{Gillett1973} who, based on 
ground observations, detected three prominent emission bands 
peaking at 8.6, 11.3 and 12.7$\mum$ in the 8--14$\mum$ spectra
of two planetary nebulae, NGC\,7027 and BD\,+\,30$\degr$3639. 
Two years later, Merrill et al.\cite{Merrill1975} reported the detection of
a broad emission band at 3.3$\mum$, again in NGC\,7027.
Also around that time, airborne observations became possible.
This led to the detection of two additional, ground-inaccessible 
intense emission bands at 6.2 and 7.7$\mum$ 
in NGC\,7027 (ref.\cite{Russell1977}) 
and M82, an external galaxy\cite{Willner1977},
%with the 4--8$\mum$ spectrophotometer 
%on board the Kuiper Airborne Observatory (KAO).
with the {\it Kuiper Airborne Observatory} (KAO).
Subsequently, all these features 
at 3.3, 6.2, 7.7, 8.6, 11.3, and 12.7$\mum$
were found to be widespread throughout the Universe
%(e.g., see Allamandola 1996).
and closely related to form a family,
exhibiting an overall similar spectral profile
among different sources 
(see Fig.\,\ref{fig:pahspec}).
%
%(e.g., see Aitken 1981). 
%

Although the exact nature of their carriers remains
unknown --- because of this, they are collectively 
known as the ``unidentified'' IR emission (UIE) features
--- the hypothesis of 
%polycyclic aromatic hydrocarbon (PAH)
PAH molecules as the carriers\cite{Leger1984,ATB1985} 
has gained widespread acceptance 
and extreme popularity.
The PAH model attributes the UIE bands to 
the vibrational modes of PAHs composed 
of fused benzene rings of several tens to
several hundreds of C atoms,
with the 3.3$\mum$ band assigned to
C--H stretching modes,
the 6.2$\mum$ and 7.7$\mum$ bands to
C--C stretching modes,
the 8.6$\mum$ band to
C--H in-plane bending modes,
and the 11.3 and 12.7$\mum$ bands to
C--H out-of-plane (CH$_{\rm oop}$) bending modes.
The relative strengths of these bands 
depend not only on the size, structure, 
and charging of the PAH molecule,
but also on the local physical 
conditions\cite{ATB1989,DL01,LD2001,DL07}.

It is now well recognized that PAHs are
an essential component of the interstellar medium (ISM)
and play an important role in many aspects 
of astrophysics. 
They account for $\simlt$15\%
of the interstellar carbon\cite{LD2001,DL07,Zubko2004,Siebenmorgen2014,Jones2017}
and their emission accounts for up to 20\% 
of the total IR power of the Milky Way 
and star-forming galaxies\cite{Smith2007,Tielens2008}.
Therefore by implication, they must be an important
absorber of starlight\cite{Joblin1992,Cecchi2008,Mulas2013} 
and are possibly related to or even responsible for 
some of the longstanding unexplained interstellar phenomena 
(e.g., the 2175$\Angstrom$ extinction 
bump\cite{Joblin1992,LD2001,Steglich2011},
the diffuse interstellar bands\cite{Salama2011},
the blue and extended red photoluminescence 
emission\cite{Witt2014},
and the ``anomalous microwave emission''\cite{Draine2003,Dickinson2018}).
%an important Galactic foreground of 
%the cosmic microwave background 
%radiation\cite{Draine2003}).
%
%PAHs could also be the source
%material for C$_{60}$\cite{Berne2012},
%
PAHs profoundly influence the thermal budget 
and chemistry of the ISM.
They dominate the heating of
the gas in the diffuse ISM 
as well as the surface layers 
of protoplanetary disks
by providing photoelectrons\cite{BT94,WD01,Kamp2004}.
%%(Lepp \& Dalgarno 1988,
%%Verstraete et al.\ 1990, 
%%
As an important sink for electrons,
PAHs dominate the ionization balance
in molecular clouds and hence they influence
the ion-molecule chemistry and the ambipolar
diffusion process that sets the stage
for star formation\cite{Verstraete2011}.

In this review, we provide an overview of
the current state of knowledge of the astrophysics
of PAHs, with details on their observational 
characteristics obtained from 
the {\it Spitzer Space Telescope}.
Special attention is also paid to 
their diagnostic capabilities to probe 
the local physical and chemical conditions
as well as their reactions to different environments.
We focus on {\it Spitzer} results,
but the science case often builds on pioneering
observations performed prior to {\it Spitzer} 
with ground-based, airborne and space telescopes, 
in particular the {\it Infrared Space Observatory} (ISO). 
%It is therefore inevitable to 
We discuss the properties 
of PAHs in the context of both {\it Spitzer} observations 
and observations obtained with prior and other 
contemporary telescope facilities.

%%% Figure 1 %%%
%%% Figure 1 %%%

% ... and with an eye to current and future surveys 
% and scientific application challenging 
% the current state of the art. 
 
\section*{PAHs in the pre-{\it Spitzer} era}
Over the intervening 30 years between the first detection
of PAHs in 1973 (ref.\cite{Gillett1973})
%by Gillett et al.\ (1973) 
and the launch of
{\it Spitzer} in 2003, 
%(Werner et al.\ 2004), 
numerous ground-based, airborne and spaceborne 
observations have substantially promoted or even
revolutionized our understanding of PAHs in astrophysics.
%Combined with laboratory measurements 
%(Hudgins \& Allamandola 1999, 2004) and 
%quantum-chemical computations (Langhoff 1996),
These observations have established that PAHs 
are an ubiquitous and abundant component of 
a wide variety of astrophysical regions, 
ranging from planetary nebulae, 
protoplanetary nebulae, reflection nebulae, 
H{\sc ii} regions, the Galactic IR cirrus, 
and protoplanetary disks around 
Herbig Ae/Be stars
to the ISM of both normal and active nearby 
galaxies\cite{Hudgins2005ASP,Tielens2008}.
%(e.g., see Hudgins \& Allamandola 2005, 
%Sauvage et al.\ 2005, Verma et al.\ 2005).
%
%
%\subsection*{Excitation mechanism}
Most notably, 
Sellgren et al.\cite{Sellgren1983}
showed that in three reflection nebulae
(NGC\,7023, NGC\,2023, and NGC\,2068)
%both 
the 3.3$\mum$ feature profile 
%and the color temperature of 
%the underlying smooth continuum emission
shows very little variation 
with distance from the central star,
%revealing that their carriers must be 
%nano-sized or smaller\cite{Sellgren1984}.
%For grains or molecules that small,
%their heat capacities are smaller than 
%or at most comparable to the energy of 
%the starlight photons that excite them\cite{Greenberg1968}.
%Upon absorption of a single stellar photon,
%they are stochastically heated to high temperatures
%to emit at $\simali$3$\mum$,
%unattainable by bulk, submicron-sized large grains. 
%Exposed to starlight of different levels of intensity 
%(e.g., at different distances from the exciting stars 
%in reflection nebulae), stochastically-heated grains
%or molecules emit in the IR essentially with the same
%spectral profile characterized by the same temperature\cite{DL01},
%while for large grains their equilibrium 
%temperatures approximately decrease with the one-third 
%power of the distance from the star\cite{Yang2016NAR}.
%
%The observations of Sellgren et al.\cite{Sellgren1983} 
%supported 
revealing the emission mechanism 
of the PAH bands as due to the IR fluorescence
from molecule-sized species vibrationally 
excited by individual UV/visible 
photons\cite{Greenberg1968,Sellgren1984,ATB1989}.
%(Allamandola et al.\ 1979, Sellgren 1984, Li 2004).

%\subsection*{Ground-based observations}
Thanks in large part to the fact that the 3$\mum$ region
and the 8--14$\mum$ region are accessible to 
ground-based telescopes, in the 1980s and 1990s
the C--H stretching bands at 3.3$\mum$ and 
the CH$_{\rm oop}$ bending bands at 11.3$\mum$ 
were the subject of extensive scrutiny. 
Careful observations revealed a great deal about 
the detailed spectral and spatial structures of 
the emission bands in these regions.
The C--H stretch near 3$\mum$ exhibits a rich spectrum: 
the dominant 3.3$\mum$ feature is usually accompanied 
by a weaker feature at 3.4$\mum$ 
along with an underlying plateau 
extending out to $\simali$3.6$\mum$.
In some objects, a series of weaker features 
at 3.46, 3.51, and 3.56$\mum$ are also seen 
superimposed on the plateau, 
showing a tendency to decrease 
in strength with increasing 
wavelength\cite{Geballe1985, JourdaindeMuizon1986, Joblin1996}.
Similarly, the 11.3$\mum$ band is also generally 
accompanied by a second distinct, but weaker, feature 
near 12.7$\mum$. Weaker features are also often present 
near 11.0, 11.5, 12.0, and 13.5$\mum$
(ref.\cite{Cohen1985}), arising from the CH$_{\rm oop}$ 
bending vibrations of PAHs.
%

%%% KAO %%%
%\subsection*{KAO}
%
In the 1970s and 1980s, the 6.2 and 7.7$\mum$ bands
were only accessible from KAO. 
KAO observations had already revealed 
significant variations in the PAH feature profiles 
among different types of sources.
Typically, harsh environments like planetary nebulae, 
reflection nebulae and H{\sc ii} regions, 
in which the dust has been heavily processed, 
show ``normal''-looking PAH spectra.
The 7.7 and 8.6$\mum$ bands of these sources
are well-separated.
In contrast, some benign protoplanetary nebulae, 
in which the dust is relatively fresh, exhibit 
a broad 8$\mum$ complex\cite{Tokunaga1997}.
The CH$_{\rm oop}$ bending bands of those heavily processed
sources also differ from that of less processed sources: 
while the former exhibit a prominent 11.3$\mum$ band 
and a weaker but distinct 12.7$\mum$ band, 
the latter intend to only show a broad complex 
at $\simali$12$\mum$.
In addition, those less processed sources 
often have a much stronger 3.4$\mum$ band
(which is often comparable to or even 
stronger than the 3.3$\mum$ band).
%
%The PAH spectral classification was later more
%thoroughly explored by Peeters et al.\ (2002)
%based on ISO data.
%
Finally, KAO observations had also showed that 
the 7.7$\mum$ band may be comprised of at least
two variable components\cite{Cohen1989}.
%

%%% ISO %%% 
%\subsection*{The ISO Legacy}
ISO provided the first opportunity to obtain 
complete IR spectra from 2.4--197$\mum$ 
unobscured by telluric absorption. 
Spanning a wavelength range of 2.4 to 45$\mum$,
the {\it Short Wavelength Spectrometer} (SWS) 
on board ISO was well suited to surveying
the PAH emission and obtaining a census of 
the PAH bands\cite{Peeters2004ASP}.
%Apart from the major bands
%of PAHs at 3.3, 6.2, 7.7, 8.6, 11.3 and 12.7$\mum$ 
%which dominate the IR emission spectra of many objects,
%the ISO/SWS observations revealed that there is also 
%a wealth of weaker, secondary features at 
%3.4, 5.25, 5.75, 6.0, 6.9, 7.5, 10.5, 11.0, 13.5, 
%and 14.2$\mum$ as well as underlying broad 
%continuum emission plateaus\cite{Peeters2004ASP}.
%Also, 
A new PAH band at 16.4$\mum$ was
discovered in the ISO/SWS spectra of 
various sources\cite{Moutou2000}.
%\cite{Moutou2000,vanKerckhoven2000}.
%
Also, the aromatic C--H and C--C bands 
at 3.3 and 6.2$\mum$ were seen in {\it absorption} 
in the ISO/SWS spectra of heavily obscured local 
sources as well as sources toward the Galactic 
center\cite{Schutte1998,Chiar2000}.
%
%With an unprecedented spectral resolving power 
%of $\lambda/\Delta\lambda$\,$\approx$\,1000--2000, 
%the ISO/SWS observations found that many of 
%the prominent bands show satellite bands 
%(e.g., the 6.2$\mum$ band is accompanied 
%by a satellite feature at 6.0$\mum$, and the 11.3$\mum$ 
%band is accompanied by a satellite feature at 11.0$\mum$),
%or break up into subfeatures (e.g., the 7.7$\mum$ band 
%is resolved to consist of two components, one at 7.6$\mum$ 
%and one at a variable wavelength from 7.8 to 8$\mum$, 
%consistent with earlier KAO observations).
%
The superb spectral resolving power of ISO/SWS
found that many of the prominent bands 
show satellite bands 
%(e.g., the 6.2 and 11.3$\mum$ bands)
or break up into subfeatures\cite{Peeters2004ASP}. 
%(e.g., the 7.7$\mum$ band). 
%
%
Moreover, the ISO/SWS observations have revealed  
systematic variations in the profiles and relative 
intensities of the PAH bands from source to source 
and spatially within sources\cite{Peeters2004ASP}. 
%The profile variations of the 7.7$\mum$ band 
%seen in the ISO/SWS spectra were even employed as
%a spectral classification scheme\cite{Peeters2002}.
%
Based primarily on the variations of the 7.7$\mum$
band seen in the ISO/SWS spectra, Peeters et al.\cite{Peeters2002}
classified the PAH sources into three classes.
%while the 7.7$\mum$ band of Class A objects 
%is dominated by the 7.6$\mum$ component,
%for Class B objects it is dominated a component
%with a highly variable peak wavelength varying
%from 7.8 to 8$\mum$. The Type B sources of Tokunaga (1997) 
%are the extreme case of Class B of Peeters et al.\ (2002).
%Class C objects show a very broad band peaking 
%at $\simali$8.2$\mum$ with a weak to absent 
%8.6$\mum$ band. 
%
%Notably, 
Furthermore,
the 5.8--11.6$\mum$ ISOPHOT low-resolution 
spectrophotometer PHT-S 
%(Lemke et al.\ 1996)
on board ISO for the first time 
detected the 6.2, 7.7, 8.6 and 11.3$\mum$ 
PAH bands in the diffuse emission 
of the Galactic disk\cite{Mattila1996}.
Although the starlight intensity in the Galactic disk 
is substantially lower than that of typical reflection nebulae 
and planetary nebulae by a factor of $\simali$100--1000,
the profiles and relative intensities of the PAH bands 
are closely similar to the ``normal''-looking PAH spectra
of reflection nebulae and planetary nebulae
(see Fig.\,\ref{fig:pah4highz}d).
At about the same time, the advent of the Japanese 
{\it Infrared Telescope in Space} (IRTS) also detected 
the 3.3--11.3$\mum$ PAH bands in the diffuse emission 
near the Galactic plane\cite{Onaka1996,Tanaka1996}.
%
%The 12$\mum$ ``cirrus'' emission detected 
%serendipitously by the broadband photometer 
%on board the {\it Infrared Astronomical Satellite} 
%(IRAS)\cite{Boulanger1988} 
%also suggests the presence of PAHs in the diffuse ISM,
%e.g., Puget et al.\cite{Puget1985} proposed that 
%the 12$\mum$ emission is largely due to the PAH bands.
%Subsequent measurements by the
%{\it Diffuse Infrared Background Experiment} (DIRBE) 
%instrument on the {\it Cosmic Background Explorer} 
%(COBE) satellite confirmed the detection of 
%the 12$\mum$ ``cirrus'' emission and detected 
%additional broadband emission at 
%3.5 and 4.9$\mum$\cite{Arendt1998}
%of which the PAH bands could be the dominant 
%contributor\cite{LD2001}.
%

\section*{The {\it Spitzer} legacy of PAH astrophysics}
Owing to its up to a factor of 100 better 
sensitivity than ISO/SWS,
%With its increased sensitivity 
%up to one order of magnitude, 
{\it Spitzer} allowed one to extend the mid-IR spectroscopy 
and imaging into new regimes that ISO could not probe.
The {\it Infrared Spectrograph} (IRS) on board 
{\it Spitzer} was capable of detecting PAH emission
in objects which were too faint for ISO
(e.g., the PAH emission at 6--9$\mum$ 
of protoplanetary disks around T Tauri stars
which had previously escaped detection by 
ISO\cite{Siebenmorgen2000} 
was unambiguously detected by
{\it Spitzer}/IRS\cite{Furlan2006,Geers2006,Seok2017};
see Fig.\,\ref{fig:pahspec}e). 
%
%Also, the 8$\mum$ broadband imaging photometer of
%the {\it Infrared Array Camera} (IRAC) on board {\it Spitzer} 
%provided a unique opportunity to explore
%the spatial distribution of PAHs 
%both in extended Galactic sources 
%and in spatially resolved external galaxies,
%as the IRAC 8$\mum$ emission is dominated 
%by the 6.2, 7.7 and 8.6$\mum$ bands of PAHs. 
%
%Compared with ISO/SWS, the main limitations of 
%{\it Spitzer}/IRS were its relatively lower spectral 
%resolution and narrower wavelength coverage. 
%With a spectral resolving power of
%$\lambda/\Delta\lambda$\,$\approx$\,50--100
%in the 5--10$\mum$ wavelength range and 
%$\lambda/\Delta\lambda$\,$\approx$\,600
%in the 10--38$\mum$ wavelength range,
%{\it Spitzer}/IRS has a limited capability 
%for one to analyze the detailed substructures 
%of the PAH band profiles. 
%Also, operating at 5--38$\mum$,  {\it Spitzer}/IRS
%%unfortunately did not cover the C--H stretching 
%%band at 3.3$\mum$ of PAHs, nevertheless,
%%all the other major PAH bands 
%%were well covered by {\it Spitzer}/IRS.
%unfortunately did not cover the PAH band at 3.3$\mum$.
%
%Nevertheless, owing to its up to a factor of 100 better 
%sensitivity than ISO/SWS, {\it Spitzer}/IRS was able to 
{\it Spitzer} was able to probe PAHs 
in much larger samples spanning much wider 
varieties of astrophysical environments,
from low-UV translucent high Galactic 
latitude clouds\cite{Ingalls2011} to UV-intense regions.
Therefore, the systematic trends and characteristics 
of PAHs as well as the exploitation of the PAH bands 
as diagnostics of the physical and chemical conditions
and processes could be determined in an unbiased manner.
Also, the high sensitivity of {\it Spitzer}/IRS enabled 
unprecedented PAH spectral mapping 
of both Galactic and extragalactic
sources\cite{Sandstrom2012,Hemachandra2015,Boersma2012,Boersma2015,Shannon2016}.
%
%(e.g., Sandstrom et al.\ 2012, 
%Hemachandra et al.\ 2015, 
%Shannon et al.\ 2015, 2016,
%Yamagishi et al.\ 2016, 
%Chastenet et al.\ 2019).
%
%Moreover, 
%the 8$\mum$ broadband imaging photometer of
%the {\it Infrared Array Camera} (IRAC) on board {\it Spitzer} 
%also provided a unique opportunity to explore
%the spatial distribution of PAHs 
%both in extended Galactic sources 
%and in spatially resolved external
%galaxies\cite{Povich2007,Shivaei2017,Wu2011},
%as the IRAC 8$\mum$ emission is dominated 
%by the 6.2, 7.7 and 8.6$\mum$ bands of PAHs. 
%
%It is mainly in these areas that,
Complemented with observations by, e.g., 
ISO and the Japanese AKARI infrared satellite, 
%and recent ground-based telescopes,
{\it Spitzer} made lasting contributions 
and substantially expanded our knowledge about
the physical and chemcial properties of PAHs
and their important role in astrophysics.
%

%%% PAHs as a spectral inventory %%%
\subsection*{The {\it Spitzer} inventory of PAH spectra.}
Compared with ISO/SWS, the main limitations of 
{\it Spitzer}/IRS were its relatively lower spectral resolution
and narrower wavelength coverage.
The spectral resolution of {\it Spitzer}/IRS 
in the 5--10$\mum$ wavelength range was lower 
than that of ISO/SWS by more than an order of magnitude.
Also, operating at 5--38$\mum$, 
{\it Spitzer}/IRS unfortunately missed 
the PAH C--H stretch at 3.3$\mum$.
%
%
%With a spectral resolving power of
%$\lambda/\Delta\lambda$\,$\approx$\,50--100
%in the 5--10$\mum$ wavelength range and 
%$\lambda/\Delta\lambda$\,$\approx$\,600
%in the 10--38$\mum$ wavelength range,
%{\it Spitzer}/IRS has a limited capability 
%for one to analyze the detailed substructures 
%of the PAH band profiles. 
%Also, operating at 5--38$\mum$,  {\it Spitzer}/IRS
%%unfortunately did not cover the C--H stretching 
%%band at 3.3$\mum$ of PAHs, nevertheless,
%%all the other major PAH bands 
%%were well covered by {\it Spitzer}/IRS.
%unfortunately did not cover the PAH band at 3.3$\mum$.
%
%
%
Nevertheless, 
despite its limited spectroscopic capabilities,
{\it Spitzer}/IRS pioneered
%clearly demonstrated its success
both in discovering new PAH bands and in showcasing
the richness and complexity of the PAH spectra, 
particularly at wavelengths longward of $\simali$10$\mum$. 
The {\it Spitzer}/IRS spectroscopy of the star-forming
ring in the spiral galaxy NGC\,7331 unambiguously
revealed, for the first time, a strong, broad emission feature 
centered at 17.1$\mum$, with a width of 
$\simali$0.96$\mum$ and an intensity 
three times as strong as the 16.4$\mum$ feature
or nearly half as strong as the ubiquitous 
11.3$\mum$ band\cite{Smith2004}.
As illustrated in Fig.\,\ref{fig:pahspec}a and \ref{fig:pahspec}g,
this prominent feature is widely seen in
the {\it Spitzer}/IRS spectra of both Galactic
and extragalactic sources\cite{Smith2007,Werner2004}.
The presence of this feature had been hinted 
by the ISO/SWS spectra of Galactic 
sources\cite{Beintema1996,van Kerckhoven2000}.
In addition,
Werner et al.\cite{Werner2004} reported discovery 
of new PAH features at 15.8 and 17.4$\mum$ 
in the {\it Spitzer}/IRS spectra of NGC\,7023,
together with a new feature at 18.9$\mum$ 
whose carrier was later identified as 
C$_{60}$\cite{Cami2010,Sellgren2010}. 
The 16.4, 17.1, and 17.4$\mum$ PAH features,
resulting from in-plane and out-of-plane ring 
bending modes of the carbon skeleton,
constitute the so-called 17$\mum$ complex.
As the strongest PAH band longward of 12$\mum$,
the 17$\mum$ complex is dominated by  
the broad 17.1$\mum$ band, with the distinct flanking
subfeatures at 16.4 and 17.4$\mum$ contributing
only $\simali$20\% of the total power 
in the complex\cite{Smith2007}.

{\it Spitzer}/IRS observations reinforced 
the richness of the PAH spectra. 
Together with ISO/SWS observations,
{\it Spitzer}/IRS observations revealed that,
apart from the major bands 
at 6.2, 7.7, 8.6, 11.3 and 12.7$\mum$ 
which dominate the mid-IR emission spectra of 
many objects, there are also 
weaker, secondary features at 
5.25, 5.7, 6.0, 6.7, 8.3, 10.5, 11.0, 12.0, 13.6, 
14.2, 15.8, 16.4, 17.4 and 17.8$\mum$ 
as well as underlying broad continuum 
emission plateaus\cite{Peeters2004ASP,Smith2007,Tielens2008}.
The CH$_{\rm oop}$ bending bands at wavelengths
longward of $\simali$10$\mum$ are of particular interest. 
They are characteristic of the edge structure of PAHs
and occur at different wavelengths, 
depending on the number of neighbouring 
hydrogen atoms on one aromatic ring: 
for neutral PAHs, isolated CH$_{\rm oop}$ band (solo-CH) 
occurs at $\simali$11.3$\mum$, 
doubly adjacent CH (duet-CH) at $\simali$12.0$\mum$, 
triply adjacent CH (trio-CH) at $\simali$12.7$\mum$, 
and quadruply adjacent CH at $\simali$13.6$\mum$
(ref.\cite{Witteborn1989,Hony2001}). 
Hudgins \& Allamandola\cite{Hudgins1999} experimentally 
showed that the aromatic CH$_{\rm oop}$ bending wavelengths 
are significantly blue-shifted upon ionization.
%
%Thanks to the sensitivity of {\it Spitzer}/IRS, 
%spectral mapping observations in these wavelengths, 
%for the first time,  became possible. The data were 
%used to quantitatively investigate the spatial variations 
%of the PAH size, structure, and ionization 
%and their responses to changing 
%environments\cite{Boersma2012,Boersma2015,Shannon2016}.
%

In the 6--9$\mum$ wavelength range, 
{\it Spitzer}/IRS observations found that 
the 7.7$\mum$ band of some carbon-rich 
post-asymptotic giant branch (post-AGB) stars 
in the Large Magellanic Cloud (LMC)\cite{Matsuura2014}
and Small Magellanic Cloud (SMC)\cite{Sloan2014}
exhibits more pronounced profile variations
than previously revealed by ISO/SWS observations
for Galactic sources. 
While the profile variations of the 7.7$\mum$ band 
seen in the ISO/SWS spectra were employed as
a spectral classification scheme\cite{Peeters2002},
the {\it Spitzer}/IRS observations led to further 
classifications of the PAH spectra\cite{Matsuura2014,Sloan2014}.
Moreover,
the subtle variations in the peak wavelength of 
the 6.2$\mum$ emission band, 
commonly attributed to
polycyclic aromatic nitrogen heterocycles
--- PAHs with one or more nitrogen atoms 
substituted into their carbon skeleton\cite{Hudgins2005PANH}
--- were recently seen in the {\it Spitzer}/IRS 
spectra of starburst galaxies\cite{Canelo2018}. 
% 
 
%%% PAHs as a spectral inventory %%%
\subsection*{2D spectral and photometric mapping 
                     of PAH emission 
                     with {\it Spitzer}.}
While ISOCAM, the camera on board ISO, had 
an imaging capability with its narrow-band filters
which cover the major PAH bands,
%in the 2.5--17$\mum$ wavelength range
%at various spatial and spectral resolutions, 
its observation was quite limited 
because of its sensitivity. 
Thanks to the much higher sensitivity
of {\it Spitzer}/IRS,
2D spectral mapping observations of PAH emission
%for the first time,  
became possible. 
{\it Spitzer}/IRS provided for the first time 
significant data that allow us to study 
the spatial distribution of PAH emission 
and quantitatively investigate the spatial variations 
of the PAH size, structure, and ionization 
and their responses to changing 
environments\cite{Boersma2012,Boersma2015,Shannon2016}.
%{\bf
Bern\'e et al.\cite{Berne2007} applied 
the so-called {\it Blind Component Separation} method
to analyze the {\it Spitzer}/IRS 2D spectral mapping
data of several Galactic objects.
They showed that the observed PAH emission 
can be decomposed into three components,
respectively arising from neutral PAHs, 
ionized PAHs, and PAH clusters or very small grains.
The spatial distributions of these components
provide useful information on their photochemical 
evolution and the local physical conditions.
%and the possible transformation 
%of very small grains to PAHs.
The viability of this method depends on
how applicable the adopted template 
emission spectra of neutral PAHs, ionized PAHs, 
and PAH clusters are to various astrophysical
regions whose PAH emission spectra are diverse. 
%}

The 8$\mum$ broadband imaging photometer of
the {\it Infrared Array Camera} (IRAC) on board {\it Spitzer} 
also provided a unique opportunity to explore
the spatial distribution of PAHs 
both in extended Galactic sources 
and in spatially resolved external galaxies,
%\cite{Povich2007,Shivaei2017,Wu2011},
as the IRAC 8$\mum$ emission is dominated 
by the 6.2, 7.7 and 8.6$\mum$ bands of PAHs. 
%{\bf
The 8$\mum$ {\it Spitzer}/IRAC mapping of
the M17 H{\sc ii} region clearly revealed 
the paucity of PAH emission 
in H{\sc ii} regions, indicating the
destruction of PAHs by extreme UV photons
within H{\sc ii} regions\cite{Povich2007}.
%}
%One of the important capabilities of {\it Spitzer} 
%was 2D mapping of the spatial distribution 
%of the PAH emission. 

%However, a number of IRAC observations 
%clearly showed that the paucity of PAH 
%emission in HII regions, e.g., Povich et al. 2007 
% ApJ, 660, 346 for M17, some of which 
%should be cited. While the paucity in HII regions 
%is not the discovery by Spitzer, I think that Spitzer 
%has indeed provided quantitative results of 
%the dependence on the radiation intensity 
%(e.g., Shivaei et al., 2017, ApJ, 837, 157, 123) 
%in addition to the metallicity dependence. 
%The metallicity dependence was also studied 
%with IRAC observations (e.g., Wu et al., 2011, 
%ApJ, 730, 111). These points should be added.

%%% Figure 2 %%%
\begin{figure} %[t!]
\centering
%\hspace{-8mm}
\includegraphics[width=8cm,height=7.5cm]{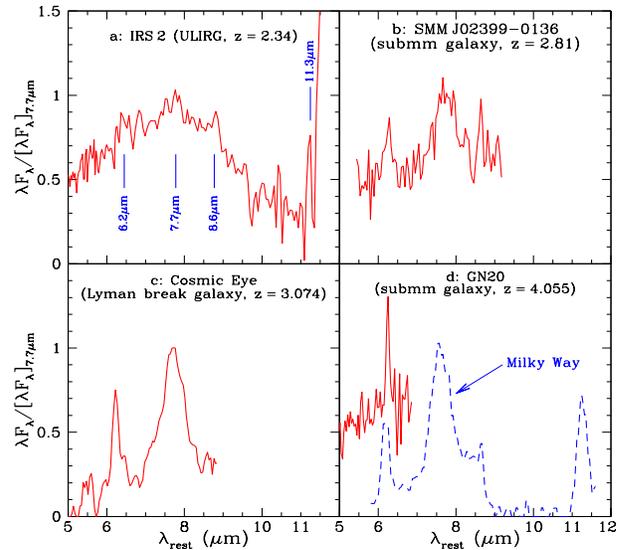}
\caption{{\bf {\it Spitzer}/IRS rest-frame spectra 
               of galaxies at high redshifts.}
               (a) ULIRG IRS\,2 ($z$\,=\,2.34)\cite{Yan2005};
               (b) Submm galaxy SMM\,J02399-0136 
                    ($z$\,=\,2.81)\cite{Lutz2005};
               (c) Lyman break galaxy Cosmic Eye 
                    ($z$\,=\,3.074)\cite{Siana2009}; and
               (d) Submm galaxy GN20 ($z$\,=\,4.055)\cite{Riechers2014}.
               Also shown in (d) is the ISOPHOT spectrum
               of the Galactic diffuse ISM
               (blue dashed line)\cite{Mattila1996}.
               }
\label{fig:pah4highz}
\end{figure}
%%% Figure 2 %%%

%%% PAHs at high z %%%
\subsection*{PAHs in the early Universe.}
Although distant galaxies are too faint to be subject 
to direct ISO/SWS spectroscopy, 
Elbaz et al.\cite{Elbaz2005}
found that the rest-frame 5--25$\mum$ 
spectral energy distributions (SEDs) 
of a sample of 16 distant luminous infrared galaxies 
(LIRGs) at redshifts $z$\,$\simali$0.1--1.2, 
%(with IR luminosity 
%$10^{11}$\,$<$\,$L_{\rm IR}/L_\odot$\,$<$\,$10^{12}$) 
constructed from the 15$\mum$ photometry of
ISOCAM  %the IR camera on board ISO, 
and the 24$\mum$ photometry of 
the {\it Multiband Imaging Photometer} (MIPS)
on board {\it  Spitzer}, indicated the presence of 
the 7.7$\mum$ PAH band 
in more than half of the sample. 
%{\it Spitzer} has clearly demonstrated 
%the power of detecting PAHs in galaxies 
%more distant than ever before.
However, as illustrated in Fig.\,\ref{fig:pah4highz},
%it was {\it Spitzer} that for the first time ever provided 
{\it Spitzer} provided the first
spectroscopic evidence for the presence of
PAHs in distant galaxies.
Yan et al.\cite{Yan2005} reported clear detection 
of multiple PAH emission bands 
in the {\it Spitzer}/IRS spectra of 
ultraluminous infrared galaxies (ULIRGs) at $z$\,$\simali$2.
%(with IR luminosity $L_{\rm IR} \simgt 10^{12}\,L_\odot$) 
%
Lutz et al.\cite{Lutz2005} also detected 
the PAH emission bands in the {\it Spitzer}/IRS 
spectra of two luminous submillimeter 
galaxies at $z$\,$\simali$2.8. 
Also, the 6.2 and 7.7$\mum$ bands of PAHs 
were clearly seen in 
the {\it Spitzer}/IRS spectrum of the Cosmic Eye, 
a strongly lensed, star-forming Lyman break galaxy 
at $z$\,=\,3.074 (ref.\cite{Siana2009}).
As shown in Fig.\,\ref{fig:pah4highz},
the PAH spectra of both the Cosmic Eye  
and the submm galaxies of Lutz et al.\cite{Lutz2005}
are ``normal''-looking, exhibiting close similarity 
to that of the Galactic diffuse ISM\cite{Mattila1996}. 
More recently,  the 6.2$\mum$ PAH band was detected 
in the {\it Spitzer}/IRS spectrum of the $z$\,=\,4.055 
submillimeter galaxy GN20, one of the intrinsically 
brightest submillimeter galaxies known\cite{Riechers2014},
%one of the most luminous dusty galaxies 
%known at high redshift. 
indicating that complex aromatic organic molecules 
were already prevalent in the young universe, 
only $\simali$1.5$\Gyr$ after the Big Bang.
As PAHs play an important role in prebiotic chemistry
which may ultimately lead to the development of 
organic life, the detection of PAHs in the early Universe 
has important astrobiological implications\cite{Bernstein1999,Kwok2016}.

Because the PAH features 
are so prevalent and intense in distant galaxies
as indicated by {\it Spitzer}/IRS observations,
in the future eras of 
the {\it James Webb Space Telescope} (JWST) 
and the {\it Space Infrared Telescope for 
Cosmology and Astrophysics} (SPICA),
they might be used as redshift indicators.
%{\bf
The integrated luminosity from the PAH features
at 6.2, 7.7, and 11.3$\mum$ has been shown to 
correlate linearly with the star formation rate (SFR)
as measured by the extinction-corrected 
H$\alpha$ luminosity for 105 galaxies
of IR luminosities 
$\simali$$10^{9}$--$10^{12}$\,$L_\odot$
%$10^{11}$\,$<$\,$L_{\rm IR}/L_\odot$\,$<$\,$10^{12}$) 
at 0\,$<$\,$z$\,$<$\,0.4 (ref.\cite{Shipley2016}). 
The luminosity of the individual 6.2, 7.7 and 11.3$\mum$
PAH bands has also been shown to 
correlate well with the IR luminosity 
for both local starburst galaxies
and high-redshift submm galaxies 
including GN20 (ref.\cite{Pope2008,Riechers2014}).
%This suggests that PAH luminosity 
%can be used as a proxy for the SFR in SMGs.
These suggest that PAH emission
could also be used as an accurate,
quantitative measure of the SFR
(but also see ref.\cite{Calzetti2007,Xie2019})
across cosmic time up to high redshifts 
to study the star formation history of the Universe.
%}
%
While the {\it Mid-Infrared Instrument} (MIRI) 
on board JWST which covers the wavelength 
range of 5 to 28$\mum$ will limit the detection
of the 6.2$\mum$ PAH band to $z$\,$\simlt$\,3, 
the SPICA {\it Far Infrared Instrument} (SAFARI), 
operating at 35--230$\mum$,
will enable the detection of PAHs
in the very first galaxies 
in the Universe\cite{Kaneda2017}.

%%% Figure 3 %%%
\begin{figure} %[t!]
\vspace{-2mm}
\centering
%\hspace{-3mm}
\includegraphics[width=8.0cm,height=7.5cm]{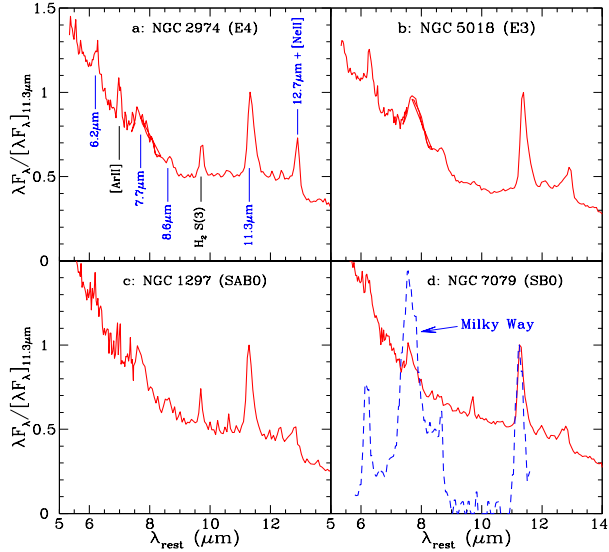}
%
%\vspace{-2mm}
\vspace{-3mm}
\caption{{\bf {\it Spitzer}/IRS spectra of 
               early-type galaxies.}
               (a) Elliptical galaxy NGC\,2974
                    of type E4 in Hubble's 
                    classification scheme\cite{Kaneda2008};
               (b) E3 elliptical galaxy NGC\,5018
                    (ref.\cite{Kaneda2008});
               (c) SAB0 lenticular galaxy NGC\,1297
                    (ref.\cite{Vega2010}); and
               (d) SB0 barred lenticular galaxy 
                    NGC\,7079 (ref.\cite{Vega2010}).
               In (a) both the PAH bands and the gaseous lines
               are labelled, where the 12.7$\mum$ band is blended
               with [Ne{\sc ii}].
               Also shown in (d) is the ISOPHOT spectrum
               of the Galactic diffuse ISM
               (blue dashed line)\cite{Mattila1996}.
               Compared to the Galactic diffuse ISM 
               of which the 6.2, 7.7 and 8.6$\mum$
               bands prevail, the PAH spectra of early-type 
               galaxies are dominated by the 11.3$\mum$ band.
               }
\label{fig:pah4ellip}
\vspace{-3mm}
\end{figure}
%%% Figure 3 %%%

%%% PAHs in elliptical galaxies %%%
\subsection*{PAHs in early-type galaxies.}
For a long time, early-type galaxies (i.e., E and S0 galaxies) 
were thought to be essentially devoid of interstellar matter.
The detection of PAHs in elliptical galaxies with 
{\it Spitzer}/IRS for the first time 
by Kaneda et al.\cite{Kaneda2005,Kaneda2008} 
demonstrated that,  
adding to the earlier detection of hot gas
% (e.g., Bregman et al.\ 1992) 
and cold dust,
% (e.g., Knapp et al.\ 1989),
elliptical galaxies contain a considerable 
amount of interstellar matter.
Vega et al.\cite{Vega2010} also reported
the detection of PAH emission
in the {\it Spitzer}/IRS spectra of S0 galaxies.
As illustrated in Fig.\,\ref{fig:pah4ellip},
the PAH spectra of early-type galaxies 
exhibit a strong enhancement 
at the 11.3$\mum$ band and substantial
suppression at the 7.7$\mum$ 
band\cite{Kaneda2005,Kaneda2008,Vega2010}.
%band\cite{Kaneda2005,Kaneda2008,Vega2010, Rampazzo2013}.
%
The peculiar 11.3/7.7 band ratios seen 
in elliptical (and S0) galaxies indicate that 
the PAH size distribution is 
%affected in the hot plasmas of ellipticals and
skewed toward appreciably larger molecules.
In the hostile environments of elliptical galaxies 
(containing hot gas of temperature $\simali$10$^7\K$), 
small PAHs can be easily destroyed 
through sputtering by plasma ions.
%impacting protons and helium nuclei.
This calls into question the origin of PAHs 
(and more generally, dust and gas) in ellipticals: 
are they from internal processes 
such as mass loss of red giants, 
or from mergers with spirals
or small dust-rich irregular galaxies,
%{\bf
or alternatively, could they be made 
in dense gas in a disk that has
cooled out of the hot ISM?
Kaneda et al.\cite{Kaneda2008} found that
the 11.3$\mum$ PAH emission of elliptical galaxies
correlates well with the dust continuum emission 
at 35$\mum$, whereas it does not correlate with 
the stellar photospheric emission at 6$\mum$. 
Therefore, they suggested that PAHs 
in elliptical galaxies are mostly of interstellar origin 
rather than of stellar origin.
However, it is also possible that,
as long as the quantities of PAHs 
and dust are both proportional to the amount of 
interstellar material in these elliptical galaxies,  
it does not matter where PAHs were formed
and one expects the dust continuum emission 
to correlate with PAH emission.
The lack of correlation between starlight
and PAH emission could also be explainable
if the excitation of PAHs in elliptical galaxies
is dominated by electronic collisions
instead of stellar photons.
%}
%
In any case,
the detection of PAHs in early-type galaxies
provide useful constraints on the evolution
of the ISM in the harsh environments 
of these galaxies.

%%% Figure 4 %%%
\begin{figure} [t!]
\centering
%\hspace{-5mm}
\includegraphics[width=8.0cm]{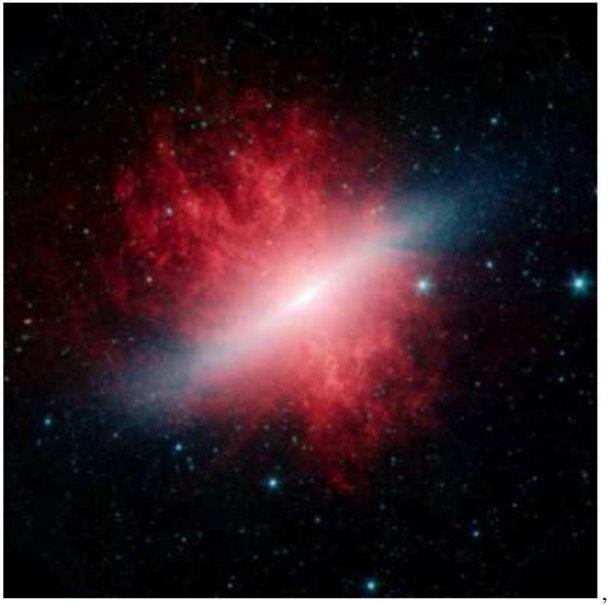},
\hspace{-3mm}
\includegraphics[width=8.0cm]{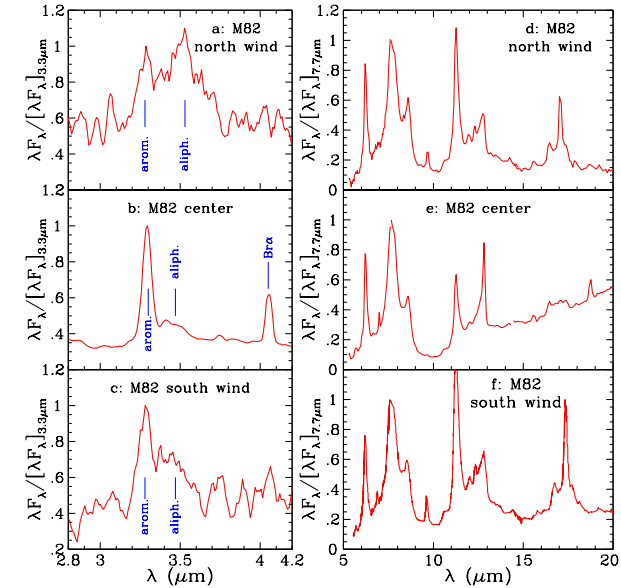}
%\subfloat[]{\includegraphics[width=5.0cm,height=5.0cm]{m82pah.eps}}
%\subfloat[]{\includegraphics[width=5.0cm,height=5.0cm]{pah4m82wind.ps}}
%\includegraphics[width=8.0cm]{pah4m82.ps}
%
\caption{{\bf PAHs in the superwind of M82.}
               Upper panel: The galaxy is shown as 
               the diffuse bar of blue light.
               The {\it Spitzer}/IRAC 8$\mum$ emission,
               dominated by PAHs, is shown in red.
               The superwind, emanating from the central
               starburst region, is evident 
               in the 8$\mum$ emission.
               The PAH emission is seen all around
               the galaxy, well beyond the cone 
               defined by the superwind. 
               %Adapted from ref.\cite{Engelbracht2006}.
               Lower panel: The near-IR AKARI and 
               mid-IR {\it Spitzer}/IRS spectra
               of the north wind (a, d), the center (b, e),
               and the south wind (c, f) of
               M82\cite{Yamagishi2012,Beirao2015}.
               %The AKARI spectra illustrates
               %the near-IR C--H stretches (b--d).
               %The {\it Spitzer}/IRS spectra show
               %the C--C stretches, C--H bending 
               %and C--C--C skeleton bending modes (e--g). 
               Most notably, while in the center 
               the 3.3$\mum$ aromatic C--H band
               is much stronger than 
               the 3.4$\mum$ aliphatic C--H band (b),
               in the wind the aliphatic band is so pronounced 
               that it even dwarfs the aromatic band (a, c).
               The intensities of the 11.3, 12.7 and 17.1$\mum$ bands 
               relatve to the 6--9$\mum$ bands
               are stronger in the wind (d, f) than in the center (f).
               %indicating that the PAH molecules in the wind
               %are somewhat larger.
               The {\it Spitzer}/IRAC 8$\mum$ image 
               is adapted from ref.\cite{Engelbracht2006}.
               }
\label{fig:m82pah}
\end{figure}
%%% Figure 4 %%%

%%% PAHs in halos and superwind %%%
\subsection*{PAHs in galaxy halos.}
Irwin \& Madden\cite{Irwin2006} analyzed the ISOCAM maps 
of NGC\,5907, an edge-on galaxy with a low SFR 
of $\simali$1.2$\Msun\yr^{-1}$,
in four narrow bands centering 
at 6.8, 7.7, 9.6, and 11.3$\mum$.
%and two broad bands centering at 6.7 and 12$\mum$.
They found that the IR emission seen in these bands
extends as far as 6.5\,kpc from the plane.
As the ISOCAM narrow-band photometry 
at 6.8, 7.7, 9.6, and 11.3$\mum$ traces 
PAH emission, they suggested that PAHs
are present in the halo of NGC\,5907.
%
%Similarly, Irwin et al.\ (2007) found the ISOCAM 
%6.7$\mum$ image of NGC\,5529, an edge-on 
%normal galaxy, exhibits a PAH-dominated mid-IR 
%halo extending to $\simali$10\,kpc from the plane. 
%

Engelbracht et al.\cite{Engelbracht2006} obtained
the {\it Spitzer}/IRAC images of the archetypal 
starburst, superwind galaxy M82
in which the central starburst drives 
a galactic outflow or superwind 
perpendicular to the galaxy plane.
As shown in Fig.\,\ref{fig:m82pah},
the IRAC 8$\mum$ emission, 
generally attributed to PAHs,
extends at least 6\,kpc from the plane
and is seen all around the galaxy,
well beyond the cone 
defined by the superwind. 
Engelbracht et al.\cite{Engelbracht2006} 
also obtained the {\it Spitzer}/IRS spectra
for M82 and clearly detected the 16.4 and 
17.1$\mum$ bands of PAHs 
in its superwind region.

Beir\~ao et al.\cite{Beirao2015} mapped 
the superwind/halo region of M82 
with {\it Spitzer}/IRS and obtained 
spatially resolved PAH spectra.
They found that the interband ratios 
of the 6.2, 7.7 and 11.3$\mum$ features  
vary by more than a factor of five 
across the superwind region.
As illustrated in Fig.\,\ref{fig:m82pah},
the intensities of the 11.3, 12.7 and 17.1$\mum$ bands 
relative to the 6.2, 7.7 and 8.6$\mum$ bands
are stronger in both the north and south superwinds 
than in the center, suggesting that the PAH molecules 
in the superwind are somewhat larger,
probably resulting from preferential destruction 
of smaller PAHs in the harsh superwind
by X-rays and/or shocks.

With AKARI, Yamagishi et al.\cite{Yamagishi2012} 
obtained the spatially-resolved 2.5--4.5$\mum$ 
near-IR spectra of M82 and detected the 3.3$\mum$ 
PAH emission as well as the 3.4$\mum$ emission
complex both in the center and in the superwind,
%located at a distance of 2\,kpc 
%away from the plane, with the intensity ratios of 
%the 3.4--3.6$\mum$ features to the 3.3$\mum$ feature 
%appreciably increasing with distance from the plane.
%
%Our results clearly confirm the presence of small PAHs 
%even in the harsh environment of the halo of M 82. 
%The results also reveal that the aliphatic hydrocarbons 
%emitting the 3.4--3.6$\mum$ features are unusually 
%abundant in the halo, suggesting that small carbonaceous 
%grains are produced by shattering of larger grains 
%in the galactic superwind.
%
% extraplanar dust in spiral galaxies...
%
%
with the intensity ratio of the 3.4$\mum$ complex 
to the 3.3$\mum$ feature 
appreciably increasing with distance from the plane.
As shown in Fig.\,\ref{fig:m82pah},
in the center the 3.4$\mum$ feature is much 
weaker than the 3.3$\mum$ feature,
in contrast, in the superwind
the 3.4$\mum$ feature supercedes the 3.3$\mum$ feature
and dominates the near-IR spectra. 
While the 3.3$\mum$ emission arises from
the aromatic C--H stretch,
the 3.4$\mum$ complex
is commonly attributed to
the aliphatic side chains attached 
as functional groups to the aromatic skeleton of
PAHs\cite{Geballe1985,JourdaindeMuizon1986,Joblin1996},
although superhydrogenated PAHs 
(i.e., PAHs whose edges contain excess
H atoms) of which the extra H atom converts
the originally aromatic ring into an aliphatic 
ring\cite{Schutte1993,Bernstein1996,Sandford2013,Steglich2013,Yang2020},
and the anharmonicity of the aromatic C--H 
stretch\cite{Barker1987, Maltseva2016}
could also be responsible for 
the 3.4$\mum$ feature.
The intensity ratio of the 3.4$\mum$ aliphatic band 
to the 3.3$\mum$ aromatic band is often 
used to derive the aliphatic fraction 
(i.e., the fraction of carbon atoms in aliphatic form) 
of PAHs\cite{LD2012,Yang2013,Yang2016NAR}.
It is puzzling that the PAH aliphatic fraction is 
considerably higher in the superwind than in the center.
One would imagine that in the harsh superwind 
environment PAHs would be easily stripped off
any aliphatic sidegroups.
For PAHs with an appreciable aliphatic fraction,
the 6.85 and 7.25$\mum$ aliphatic C--H 
deformation bands\cite{Yang2016MN}
would show up (e.g., see Fig.\,\ref{fig:pahspec}e).
However, a close inspection of Fig.\,\ref{fig:m82pah}
reveals no evidence for these two bands
in the {\it Spitzer}/IRS spectra of M82,
neither in the center nor in the superwind.
It is also puzzling how small PAHs of 
$\simali$20--30 C atoms which emit
at 3.3$\mum$\cite{DL07} could survive
in the superwind. 
Yamagishi et al.\cite{Yamagishi2012}
suggested that they are produced 
{\it in situ} by shattering of large grains. 
Alternatively, they may be protected 
in clumps\cite{Micelotta2010}.

Finally, the {\it Spitzer}/IRS observations of 
the Galactic bulge also detected PAHs
in a local Galactic wind environment\cite{Shannon2018}. 
The presence of PAHs in the diffuse ionized 
halo of NGC\,891, an edge-on spiral, was also
spectroscopically revealed by {\it Spitzer}/IRS
through the observations of the PAH bands 
at 11.3, 12.7, and 16.4$\mum$\cite{Rand2008}.
The detection of PAHs in galaxy halos 
or superwinds provides valuable insight 
not only into the destruction and survival 
of PAHs in hostile environments, but also 
the mechanism for transporting PAHs 
from the plane to the halo 
through supernova explosion, 
stellar winds, 
and radiation pressure.

%
%McCormick et al.\ (2013) used the {\it Spitzer}/IRAC images 
%to investigate PAH emission from a sample of 16 local galaxies 
%with known winds. 
%... has revealed detailed PAH structure in the winds 
%and allowed us to measure extraplanar PAH emission. 
%They identified extraplanar PAH features on scales of 
%$\simali$0.8--6.0\,kpc and found a nearly linear correlation 
%between the amount of extraplanar PAH emission 
%and the total infrared flux, a proxy for star formation 
%activity in the disk. Their results also indicate a correlation 
%between the height of extraplanar PAH emission and 
%star formation rate surface density, which supports the idea 
%of a surface density threshold on the energy or momentum 
%injection rate for producing detectable extraplanar wind material. 
%

%The detection of PAHs both in the halo of
%the low SFR galaxy NGC\,5907 and 
%in the superwind of the starburst galaxy M82
%provides valuable insight not only into 
%the destruction and survival of PAHs 
%in hostile environments but also 
%the dynamics of PAHs, e.g., the mechanism   
%for transporting PAHs from the plane to the halo.
%While PAHs could get ``levitated'' to high latitudes
%in M82 by mechanical energies from supernova 
%explosion and winds, the outward expulsion
%of PAHs in NGC\,5907 is more likely caused by
%radiation pressure.

\subsection*{PAHs in galaxy mergers.}
Galaxy mergers, 
%the most violent type of galaxy interaction
occuring when two (or more) galaxies collide, 
could trigger starbursts and lead to the formation 
of tidal tails stretching $\simgt$100\,kpc 
from the site of the collision\cite{Sanders1996}.
Higdon et al.\cite{Higdon2006} 
obtained the {\it Spitzer}/IRS spectra of 
two faint tidal dwarf galaxies, 
NGC\,5291\,N and NGC\,5291\,S, and detected 
PAH emission bands at 6.2, 7.7, 8.6, 11.3, 12.7, 
and 16.4$\mum$ which are remarkably similar 
%to those of starburst galaxies. 
%and NGC\,7023 (reflection nebula).
to those of normal star forming galaxies.
In contrast, Haan et al.\cite{Haan2011} 
mapped the PAH bands with {\it Spitzer}/IRS
in eight major merger systems of the Toomre Sequence 
%          ({\small NGC}\,4676, {\small NGC}\,7592, 
%           {\small NGC}\,6621, {\small NGC}\,2623, 
%           {\small NGC}\,6240, {\small NGC}\,520, 
%           {\small NGC}\,3921, and {\small NGC}\,7252).
and found that the spatially resolved 6.2/7.7 and 11.3/7.7
interband ratios 
%of many regions 
are often too large to be explainable 
by the canonical PAH model\cite{DL01}.
Murata et al.\cite{Murata2017} 
examined the relationship 
of the PAH emission spectra obtained with AKARI
with galaxy merger in 55 star-forming galaxies 
at $z$\,$<$\,0.2. They found that PAHs are relatively
underabundant in merger galaxies than non-merger 
galaxies and suggested that PAHs are partly destroyed 
by the intense UV radiation and large-scale shocks
during merging processes of galaxies. 
Based on {\it Spitzer}/IRS and AKARI data,
Onaka et al.\cite{Onaka2018} detected PAH emission
and found that small grains are deficient
in the tidal tails of two galaxy mergers, 
NGC\,2782 (Arp\,215) and NGC\,7727 (Arp\,222).
They suggested that PAHs are formed 
from the fragmentation of small grains  
during merger events. 
%
%
%V\"ais\"anen et al.\ (2012):
%The nuclear PAH emission of 
%merger system NGC 1614: 
%rings within rings 
%-- the inner kpc of the luminous infrared galaxy NGC 1614 -- 
%A broad ring of 3.3$\mum$ PAH emission is found at 
%a distance of approximately 200 pc from the core. 

%%% Figure 5 %%%
\begin{figure*}%[t!]
\centering
%\hspace{-5mm}
\vspace{-7mm}
\includegraphics[width=9cm,angle=90]{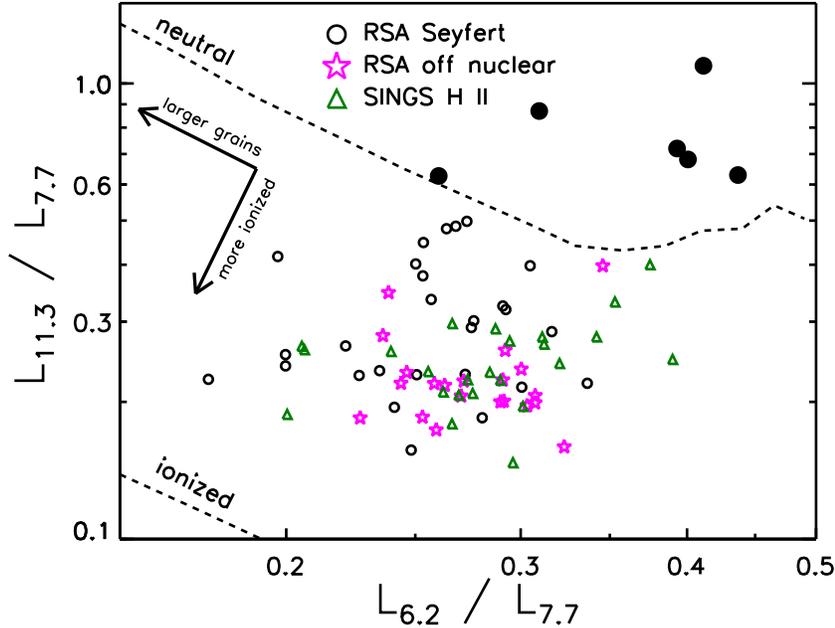}
\caption{{\bf Relative strengths of the 6.2, 7.7, and 11.3$\mum$
              PAH features for Seyfert nuclei 
              (open or filled circles), 
              off-nuclear regions (stars), 
              and H{\sc ii} galaxies (triangles).}
              The Seyfert galaxies are from the {\it Revised
              Shapley-Ames} (RAS) catalog of bright galaxies.
              $L_{6.2}$, $L_{7.7}$ and $L{11.3}$ are repsectively
              the power emitted from the 6.2, 7.7 and 11.3$\mum$ features.
              %\bf Relative strengths of the 6.2, 7.7, and 11.3$\mum$
              %PAH features for RSA Seyfert nuclei, 
              %off-nuclear regions, and SINGS H{\sc ii} galaxies.}
              The dashed lines correspond to predictions 
               of Draine \& Li\cite{DL2001}
              for completely neutral and completely ionized PAHs
              %of varying sizes; 
              with their sizes increasing from right to left;
              the permitted region of the diagram is bounded 
              by these two lines.
              The Seyferts highlighted as filled circles all lie beyond
              the range of model predictions, even for completely 
              neutral molecules.
              It is likely that heavy processing causes PAHs to
              have an open, irregular structure and hence a higher 
              H/C ratio and consequently a higher 11.3/7.7 ratio.
              Adapted from ref.\cite{Diamond2010}.
               }
\label{fig:pah4agn}
\end{figure*}
%%% Figure 5 %%%

\subsection*{PAHs in active galactic nuclei.}
In the harsh environment around active galactic nuclei 
(AGNs) --- rich in extreme UV and soft X-ray photons --- 
PAHs would be destroyed and PAH emission is
not expected to be present\cite{Voit1992,Siebenmorgen2004},
as first noticed by Roche et al.\cite{Roche1991}
in the ground-based mid-IR spectra of
the nuclear regions of active galaxies.
Genzel et al.\cite{Genzel1998} proposed to
use the strength of the 7.7$\mum$ PAH feature 
as a discriminator of starburst and AGN activity 
in ULIRGs, with an aim to determine
whether an ULIRG is powered by recently 
formed massive stars or by a central AGN. 
However, PAH emission was reportedly detected 
within $\simgt$10$\pc$ of AGNs,
suggesting that PAHs could survive in close
proximity to AGNs and be excited by photons
from AGNs\cite{Esquej2014,Jensen2017}.
%
%{\bf
Based on the {\it Spitzer}/IRS spectra of 91 
Seyfert galaxies, Tommasin et al.\cite{Tommasin2010} 
also found that, going from AGN-dominated 
to starburst-dominated objects,
the 11.3$\mum$ PAH feature  
remains almost constant in flux,
although its equivalent width increases.
They argued that PAHs could survive 
in the highly ionized medium of AGNs, 
while the PAH feature appears weaker 
in the most powerful ones because it is masked 
by the strong underlying AGN continuum
(see their Fig.\,14).
%}
%
On the other hand,
AKARI and {\it Spitzer}/IRS observations 
have shown that PAHs could be destroyed 
by the intense, hard radiation in starburst galaxies 
and in the H{\sc ii} regions within star-forming 
galaxies\cite{Murata2014,Maragkoudakis2018}.
Therefore, caution should be taken when one uses 
the PAH emission as a star formation tracer 
within a kpc around AGNs or for intense starbursts.

Nevertheless,
numerous observations of Seyferts and LINERs 
with {\it Spitzer}/IRS have clearly shown that 
the PAH emission significantly weakens 
in AGN-hosting galaxy nuclei
and the PAH spectra differ considerably
from that of star-forming galaxies.
More specifically, 
%as illustrated in Fig.\,\ref{fig:pah4agn},
the PAH bands at 6.2, 7.7, and 8.6$\mum$ 
of Seyferts and LINERs 
are often substantially suppressed 
relative to the 11.3$\mum$ 
band\cite{Smith2007, ODowd2009, Diamond2010, Wu2010}.
These trends have been interpreted as
the preferential destruction of small PAHs 
by the hard radiation field of AGNs.

As demonstrated in Fig.\,\ref{fig:pah4agn}, 
the relative strengths of the PAH bands
provide powerful diagnostics of the physical 
and chemical properties of the PAH molecules 
(e.g., their sizes, charging, and structural characteristics).
%themselves as well as the local physical conditions
%(e.g., the starlight intensity $U$ and hardness, 
%gas temperature $\Tgas$
%and electron density $n_e$). 
%
%
%
Both laboratory measurements and quantum-chemical computations
have shown that the 3.3 and 11.3$\mum$ features
arise primarily from neutral PAHs, 
while the 6.2, 7.7, and 8.6$\mum$ features are dominated
by the emission of ionized 
PAHs\cite{Allamandola1999,Hudgins2005ASP,Bauschlicher2018}.
%Boersma et al.\ 2014, Bauschlicher et al.\ 2018).
%(e.g., see  Langhoff 1996, Bauschlicher \& Langhoff 1997,
%Hudgins \& Allamandola 1997, 1999, 2004,
%Hudgins \& Sandford 1998a,b, Ellinger et al.\ 1998, 
%Allamandola et al.\ 1999, Bauschlicher 2002, 2008, 2009,
%Kim \& Saykally 2002, Pauzat \& Ellinger 2002,
%Mulas et al.\ 2006, Malloci et al.\ 2007, 
%Ricca et al.\ 2010, 2012, Candian et al.\ 2014, 
%Candian \& Sarre 2015).
%
%
%Both laboratory measurements 
%and quantum-chemical computations
%have also shown that, 
%Upon insertion of a nitrogen atom,
%the 6.2 and 7.7$\mum$ C--C stretches 
%and the 8.6$\mum$ C--H in-plane bending
%of both neutral and cationic PAHs 
%exhibit a two-fold increase in
%intensity\cite{Mattioda2003,Mattioda2017}.
%
Meanwhile, whether a PAH molecule will be ionized
or neutral is controlled by the starlight intensity, 
electron density and gas temperature\cite{BT94,WD01}.
Therefore, the band ratios involving the neutrals and ions,
such as the 7.7/11.3 ratio, are useful tools for probing 
the charging of the emitting molecules 
and hence for probing the local physical 
conditions\cite{DL01,Galliano2008BandRatio}.

By comparing the observed PAH interband ratios
with the model expectations for 
neutral and ionized PAHs,
one could determine the PAH size
and ionization fraction 
(i.e., the probability of finding 
a PAH molecule 
in a nonzero charge state)\cite{DL01,Galliano2008BandRatio}.
However, as shown in Fig.\,\ref{fig:pah4agn},
Diamond-Stanic \& Rieke\cite{Diamond2010}
found that the 11.3/7.7 ratios of
a number of Seyferts lie beyond 
the range of model predictions, 
even for completely neutral PAHs.
%Similar extreme 11.3/7.7 ratios were 
%also seen in the compact H{\sc ii} region 
%IRAS\,12063-6259\cite{Stock2013}
%which the PAH model was unable to reproduce 
%%the observed band ratios 
%even with completely neutral grains.
While large 11.3/7.7 ratios could be produced 
by large neutral PAHs, they would be expected 
to have small 6.2/7.7 ratios, 
inconsistent with the observations.
This is because, for a given ionization, 
larger PAHs tend to emit more at longer wavelengths
(i.e., larger 11.3/7.7 ratios 
and smaller 6.2/7.7 ratios)\cite{DL01}.  
%
%{\bf
The discrepancy between the observed 
and model-predicted 11.3/7.7 ratios 
cannot be resolved by nitrogen-containing PAHs.
%since they would produce even lower 11.3/7.7 ratios.
Upon insertion of a nitrogen atom,
the 6.2 and 7.7$\mum$ C--C stretches 
and the 8.6$\mum$ C--H in-plane bending
of both neutral and cationic PAHs 
exhibit a two-fold increase in
intensity\cite{Mattioda2003,Mattioda2017}.
Therefore, nitrogen-containing PAHs
would produce even lower 11.3/7.7 ratios.
%}
%
We suggest that these extreme 11.3/7.7 ratios
could arise from catacondensed PAHs with
an open, irregular structure. 
Catacondensed PAHs have more H atoms 
(on a per C atom basis)
than compact, pericondensed PAHs
(e.g., coronene, ovalene, 
circumcoronene)\cite{Hony2001,Shannon2016}
and hence their emission spectra would
have a higher 11.3/7.7 ratio.
While the model tracks shown in Fig.\,\ref{fig:pah4agn}  
for ionized and neutral PAHs at varying sizes 
were calculated from pericondensed PAHs\cite{DL01},
Heavy processing 
in the nuclear regions of active galaxies 
may cause PAHs to have an open, irregular structure
and thus a higher 11.3/7.7 ratio.
Alternatively, the extreme 11.3/7.7 ratios could 
result from an inappropriate subtraction of 
the continuum emission underlying the PAH bands. 
%

%%% Figure 6 %%%
\begin{figure} %[t!]
\centering
%\hspace{-5mm}
\includegraphics[width=8cm]{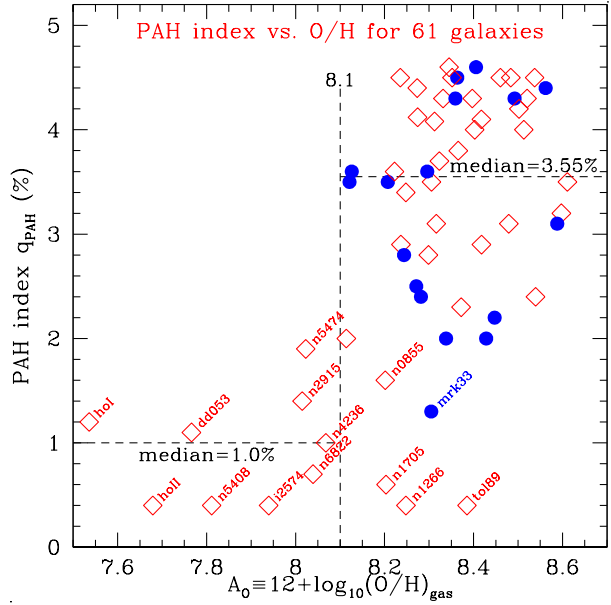}
\caption{{\bf PAH abundance vs. galaxy metallicity.}
              The galaxy metallicity 
              $A_{\rm O} \equiv 12+\log({\rm O/H})$
               is measured 
              by the gas-phase oxygen abundance 
              (relative to H).
              The PAH abundance is measured by 
               the mass fraction of dust in PAHs,
               with an uncertainty of $\simali$0.1\%.
               Low metallicity galaxies are always 
               deficient in PAHs.
               Filled circles are for galaxies having 
               submillimter data; diamonds are for
	       galaxies lacking submillimeter data.
               Adapted from ref.\cite{Draine2007}.
               }
\label{fig:pah4metalpr}
\end{figure}
%%% Figure 6 %%%

\subsection*{PAHs in low-metallicity galaxies.}
The deficiency or lack of PAHs in low-metallicity galaxies
was recognized in the pre-{\it Spitzer} era.
%The Small Magellanic Cloud (SMC), an irregular dwarf galaxy
%with a metallicity ($Z$) just $\simali$1/10 of solar ($Z_\odot$), 
%exhibits a local minimum at $\simali$12$\mum$ 
%in its IR emission spectrum, 
%suggesting the lack of PAHs (Li \& Draine 2002).
%
The $\simali$5--16$\mum$ mid-IR spectrum of 
SBS\,0335-052, a blue compact dwarf galaxy 
with an extremely low metallicity of 
$Z/Z_\odot$\,$\simali$1/41,
obtained by Thuan et al.\cite{Thuan1999} with
the {\it Circular Variable Filter} (CVF) facility 
of ISOCAM shows no evidence of PAHs,
as later confirmed by the {\it Spitzer}/IRS 
observations\cite{Houck2004}.
Madden et al.\cite{Madden2006} performed a more systematic 
investigation of the ISOCAM CVF mid-IR spectra of
a large number of metal-poor dwarf galaxies and 
found that the PAH features are substantially suppressed.
The advent of {\it Spitzer} not only confirmed 
the paucity of PAH emission in low-metallicity 
environments, but more importantly, also established
a trend of decreasing PAH emission with metallicity 
and a threshold in metallicity
below which the PAH abundance drops drastically.

The {\it Spitzer}/IRS spectra of a large number of
metal-poor blue compact dwarf galaxies
with metallicities from $Z/Z_\odot$\,$\simali$0.02
to $\simali$0.6 show much weaker PAH features 
than typical starburst galaxies,
with a substantial weakening at
$Z/Z_\odot$\,$\simlt$\,0.2 or
$12$\,+\,$\log({\rm O/H})$\,$\sim$\,7.9
(ref.\cite{Wu2006,Hunt2010}).
%Hunt et al.\ 2005.
%
Engelbracht et al.\cite{Engelbracht2005} examined 
the IRAC 8$\mum$ and MIPS 24$\mum$ emission
of 34 galaxies spanning two decades in metallicity.
They found that $f_\nu(8\mu {\rm m})$/$f_\nu(24\mu {\rm m})$ 
--- the ratio of the 8$\mum$ emission
to the 24$\mum$ emission ---
%$f_\nu(8\mu {\rm m})$/$f_\nu(24\mu {\rm m})$
drops abruptly from 
%$f_\nu(8\mu {\rm m})$/$f_\nu(24\mu {\rm m})$
$\simali$0.7
for galaxies with $Z/Z_\odot$\,$>$\,1/3 to 
%$f_\nu(8\mu {\rm m})$/$f_\nu(24\mu {\rm m})$
$\simali$0.08
for galaxies with $Z/Z_\odot$\,$<$\,1/5--1/3.
They attributed this drop to a sharp decrease 
in the 7.7$\mum$ PAH feature at 
a threshold metallicity of
$12$\,+\,$\log({\rm O/H})$\,$\sim$\,8.2.
By modeling the PAH and dust IR emission
of 61 galaxies from the {\it Spitzer Infrared Nearby
Galaxies Survey} (SINGS),
Draine et al.\cite{Draine2007} also found that 
the PAH abundance shows a sharp change
around a threshold in metallicity of
$12$\,+\,$\log({\rm O/H})$\,$\sim$\,8.1
(see Fig.\,\ref{fig:pah4metalpr}),
corresponding to $Z/Z_\odot$\,$\simlt$\,0.23
if we take $Z/Z_\odot$\,=\,1
for $12$\,+\,$\log({\rm O/H})$\,=\,8.73
%i.e., $\left({\rm O/H}\right)_\odot$\,=\,5.37\,$\times$\,$10^{-4}$
(ref.\cite{Asplund2009}).

%In the DL07 dust model, 
%the PAH fraction is defined as the fraction 
%of the total dust mass in grains with less than 
%10$^3$ carbon atoms, and is hereafter labeled 
%$q_{\rm \sc PAH}$. 
%The Galactic diffuse neutral medium 
%$q_{\rm \sc PAH}$ lies around 4.6\%.

%There are several hypotheses for the dearth of PAHs 
%in low-metallicity environments. 
%PAHs could be exposed to more intense 
%and/or harder far-UV radiation fields 
%due to the overall decrease in dust shielding, 
%and suffer from a more efficient selective
%photo-destruction \citep[e.g.][]{Madden06}. 
%PAHs could also form in the dense ISM \citep[][]{Zhukovska16}, 
%and this process could be less effective at lower metallicities. 
%Or the low PAH abundance could be the sign of 
%a lower efficiency of forming PAH-like dust, 
%due to particular stellar evolution 
%at low metallicity \citep[][]{Galliano08}.

With {\it Spitzer}/IRS,
Gordon et al.\cite{Gordon2008} investigated 
the spatially-resolved PAH features 
in the face-on spiral galaxy M101 
which has one of the largest metallicity gradients. 
%known with $12+\log({\rm O/H})$ ranging from 
%$\simali$8.8 in the nucleus to $\simali$7.4 
%in an H{\sc ii} region $\simali$41\,kpc from the center. 
They found that the variation of the strengths of 
the PAH features correlates better with 
the radiation field hardness than metallicity. 
%as measured by [Ne iii]/[Ne ii] and [S iv/S iii].
%{\bf
On the contrary, Wu et al.\cite{Wu2011} found
that the PAH emission of 29 faint dwarf galaxies
as traced by the {\it Spitzer}/IRAC [3.6]$-$[8] color
appears to depend more directly on metallicity
than radiation hardness.
%}
%

The exact reason for the deficiency or lack of PAHs
in low-metallicity galaxies is not clear.
It is generally interpreted as more rapid destruction 
of PAHs by the more intense and harder UV radiation
(as indicated by the fine-structure line ratio of 
[Ne{\sc iii}]/[Ne{\sc ii}])
%Thornley et al.\ 2000)
in an ISM with reduced shielding by dust.
%(e.g., see Thuan et al.\ 1999, Plante \& Sauvage 2002,
%Houck et al.\ 2004, Engelbracht et al.\ 2005, 
%Madden et al.\ 2005, Wu et al.\ 2005, Draine et al.\ 2007).
%
Low-metallicity environments lack sufficient dust grains 
to shield PAHs from photodissociation by UV radiation,
by analogy with the mechanism commonly invoked to 
explain the deficit of CO emission 
in low-metallicity galaxies\cite{Xie2019}.
It could also be due to more effective destruction of PAHs 
by thermal sputtering in shock-heated gas that cools more
slowly because of the reduced metallicity.
The reduced PAH abundance in these galaxies 
might also be due to a deficiency of PAH-producing 
carbon stars and C-rich planetary nebulae,
or a suppressed formation and growth of PAHs 
in the ISM with low gas-phase C abundances\cite{Draine2007}.
%(Draine et al.\ 2007).
%
%Hogg et al.\ (2005) argued that the observed lack
%of PAHs is closely related to the low luminosity nature 
%of these galaxies.
%
%Studies have found evidence of changes 
%in PAH properties as a function of galaxy properties, 
%particularly a PAH deficiency at low metallicity  
%\citep[][]{Engelbracht05, Madden06, Draine07, 
%Galliano08, Sandstrom10, Paradis11, RemyRuyer15}. 
%
%{\bf
Seok et al.\cite{Seok2014} related 
the PAH-metallicity dependence to 
the formation mechanism of PAHs in the ISM.
They suggested that interstellar PAHs are formed 
from the shattering of carbonaceous dust grains 
in interstellar turbulence. 
Based on an exploration of the evolution 
of PAH abundance on a galaxy-evolution time-scale,   
they found that the formation of PAHs becomes 
accelerated above certain metallicity 
where shattering becomes efficient. 
%}

%Alternatively, 
The dearth of PAHs in low-metallicity 
galaxies, if they are truly young, could also result 
from a delayed production of PAHs by low-mass stars,
%injection of carbon molecules from 
%low-mass stars into the ISM,
i.e., PAH-producing carbon stars have not yet
evolved off the Main Sequence\cite{Galliano2008AGB}. 
%so that PAHs have simply 
%not had time to form 
%(Dwek 2004).
%(Dwek 2004, Galliano et al.\ 2008).
This scenario is supported by 
the metallicity-dependence of 
the 7.7$\mum$ PAH emission
of 476 distant galaxies 
spanning a wide range in metallicity 
at redshifts $1.37\simlt z\simlt2.61$.
%Compared to older galaxies, 
The PAH emission
(relative to the total IR emission)
is indeed significantly lower 
in the youngest quartile of the sample 
with an age of $\simlt500\Myr$ (ref.\cite{Shivaei2017}).
%which may be a result of the delayed production 
%of PAHs by AGB stars.
However, based on a {\it Spitzer}/IRAC 8$\mum$
imaging survey of 15 local group dwarf galaxies,
Jackson et al.\cite{Jackson2006} found that,
although these galaxies are all deficient in PAH emission, 
they have formed the bulk of their stars more than 2\,Gyr ago. 
Therefore, they argued that the paucity of PAH emission 
in dwarf galaxies is unlikely due to the shortage of AGB stars.

%%%% Box 1: major contributions %%%%
\begin{figure}
\noindent\fcolorbox{white}{colorbox}{%
%\begin{minipage}{0.5\textwidth}
%\centerline{\psfig{figure=box1.ps,width=\textwidth,angle=0}}
%\end{minipage}
\begin{minipage}{0.49\textwidth}
\fontfamily{phv}
\selectfont{Box~1: Major contributions 
   to PAH astrophysics made by 
   {\it Spitzer} observations.
   }
%\vspace{0.1cm}
\noindent 
{\small
\begin{itemize}
\item {\it Spitzer}/IRS observations have provided 
          a complete and unbiased spectroscopic census
          of the PAH emission bands in the wavelength 
          range of $\simali$5--20$\mum$,
          including the discovery of 
          a prominent emission feature 
          centered at 17.1$\mum$ 
          whose intensity is three times as strong as
          the 16.4$\mum$ feature or nearly half as strong 
          as the 11.3$\mum$ band
          (see Fig.\,\ref{fig:pahspec}a,c)\cite{Smith2004}.
\item {\it Spitzer}/IRS observations have detected
          PAH emission in external galaxies out to
          high redshifts $z$\,$\simgt$\,4, indicating 
          the prevalence of complex aromatic organic molecules 
          in the young universe, 
          only $\simali$1.5$\Gyr$ 
          after the Big Bang\cite{Riechers2014}.
\item With its high sensitivity, {\it Spitzer} has been able 
          to probe PAH emission in much larger samples 
          spanning much wider varieties of 
          astrophysical regions than ever before,
          including such hostile environments as
          AGNs\cite{ODowd2009, Diamond2010, Wu2010},
          elliptical galaxies\cite{Kaneda2005},
          galactic outflows or 
          superwinds\cite{Engelbracht2006,Rand2008,Beirao2015}, 
          galaxy mergers\cite{Higdon2006}, 
          low-metallicity 
          galaxies\cite{Houck2004,Engelbracht2005,Wu2006,Hunt2010,Draine2007},
          and supernova remnants\cite{Tappe2006,Andersen2011,Seok2012}.
          In these harsh environments, 
          the PAH spectra are often substantially
          suppressed in the 6.2, 7.7 and 8.6$\mum$ bands
          and enhanced in the 11.3$\mum$ band.
          The 6--9$\mum$ PAH emission of
          protoplanetary disks around T Tauri stars
          has also been detected, for the first time,
          by {\it Spitzer}/IRS\cite{Furlan2006,Geers2006,Seok2017}. 
\item The high sensitivity of {\it Spitzer}/IRS has enabled 
          unprecedented PAH spectral mapping 
          of both Galactic and extragalactic sources,
          providing valuable information on
          the spatial variations of the PAH characteristics 
          (e.g., size, charge, structure) and their responses
          to the changing 
          environments\cite{Sandstrom2012,Hemachandra2015,Boersma2012,Boersma2015,Shannon2016}.
\item {\it Spitzer} observations firmly established
          the PAH-metallicity relation for galaxies
          spanning over two decades in metallicity
          and revealed a metallicity threshold 
          of $12$\,+\,$\log({\rm O/H})$\,$\sim$\,8.1
          below which the PAH abundance drops 
          drastically\cite{Engelbracht2005,Wu2006,Hunt2010,Draine2007}.
\end{itemize}
}
\end{minipage}
}%
\end{figure}

\subsection*{PAHs in supernova remnants.}
The PAH spectra of supernova remnants (SNRs) 
and their implied PAH sizes contain useful
information on the shock-processing of PAHs.
With {\it Spitzer}/IRS,
Tappe et al.\cite{Tappe2006} 
performed spectroscopic observations of N132D, 
a young SNR with an age of $\simali$2500$\yr$ 
in the LMC. 
They reported the detection in N132D of
a weak PAH feature at 11.3$\mum$ and a prominent, 
broad hump at 15--20$\mum$ attributed to 
the C-C-C skeleton bending modes of large PAHs
of $\simali$4000 C atoms.
This was the first ever detection of PAHs in a SNR.
The lack of PAH features at 6--9$\mum$ and
the large ratio of the 15--20$\mum$ emission 
to the 11.3$\mum$ emission indicate that
small PAHs could have been rapidly destroyed 
in the supernova blast wave via thermal sputtering
and only larger ones have survived 
in the shocked environment.

In contrast, with AKARI and {\it Spitzer}/IRS,
Seok et al.\cite{Seok2012} 
detected the 3.3, 6.2, 7.7 and 11.3$\mum$
PAH features in N49, a middle-aged SNR ($\simali$6600\,yr)
in the LMC, and found that the 6.2/11.3 and 7.7/11.3 band 
ratios imply a predominance of small PAHs.
They suggested that,
unlike N132D,
%of which small PAHs
%were preferentially destroyed 
%and only large PAHs survived,
the PAHs in N49 are possibly 
associated with the ambient molecular cloud
(with which N49 is interacting) 
where small PAHs could survive from 
the shock\cite{Micelotta2010}.
%

%{\bf
The detetction of PAH emission 
in Galactic SNRs has also been reported 
with {\it Spitzer}/IRS\cite{Andersen2011}.
Their $\simali$5--14$\mum$ PAH spectra 
closely resemble that of Galactic 
photodissociated regions (PDRs) 
and star-forming galaxies, but considerably 
differ from that of N132D in the LMC. 
Andersen et al.\cite{Andersen2011} attributed 
the spectral difference between Galactic SNRs 
and N132D to their different environments.
While many of these Galactic SNRs are in a dense 
molecular cloud environment with a slow shock, 
N132D, in a less dense environment, has a strong 
shock which has effectively destroyed small PAHs. 
\begin{figure*}
\noindent\fcolorbox{white}{colorbox}{%
%\begin{minipage}{0.5\textwidth}
%\centerline{\psfig{figure=box1.ps,width=\textwidth,angle=0}}
%\end{minipage}
\begin{minipage}{0.98\textwidth}
\fontfamily{phv}
\selectfont{Box~2: Puzzling questions about PAHs
   raised partly from {\it Spitzer} observations.
   }
%\vspace{0.1cm}
\noindent 
{\small
\begin{itemize}
\item No exact identification of the UIE band carriers 
          has been made yet\cite{KZ2011}.
          No specific PAH molecule 
          has been identified
          in the interstellar or circumstellar space.
          %Not a single specific PAH molecule 
          %has yet been identified
          %in the interstellar and circumstellar 
          %medium\cite{KZ2011}.
% 
\item Are interstellar PAHs made in the ISM 
          or condensed in carbon star outflows 
          and subsequently injected into the ISM?
\item How do PAHs quantitatively reveal 
          the local physical and chemical conditions
          and respond to intense, extreme UV 
          and X-ray photons and shocks?
          How would the chemical structure
          (e.g., catacondensed vs. pericondensed,
           aliphaticity vs. aromaticity,
           dehydrogenation vs. superhydrogenation)
          of PAHs be affected by photo- and
          shock-processing? 
          How do the extreme 11.3/7.7 band ratios
          seen the {\it Spitzer}/IRS spectra of AGNs  
          (see Fig.\,\ref{fig:pah4agn})
          and other harsh environments
          reflect the size, charging, and structure of PAHs
          and the processing they have experienced?  
\item How accurate are PAHs as a SFR indicator? 
          While PAHs are seen in regions within
          as close as $\simgt$\,10$\pc$ of 
          AGNs\cite{Esquej2014,Jensen2017},
          {\it Spitzer}/IRS observations have shown 
          that PAHs can be destroyed by intense 
          starbursts\cite{Murata2014,Maragkoudakis2018}.
          Also, PAH emission does not
          necessarily trace exclusively young stars
          as PAHs can also be excited by visible 
          photons\cite{Uchida1998,LD2002,Mattioda2005}.
\item What is the nature of the continuum
          underlying the discrete PAH bands? 
          What is the most appropriate way to
          subtract the continuum to measure 
          the PAH emission (which is crucial
          for accurately determining the SFRs
          and the PAH band ratios)?
\item Are PAHs related to other unexplained interstellar 
          phenomena (e.g., the 2175$\Angstrom$ extinction bump, 
          the diffuse interstellar bands, 
          the blue and extended red photoluminescence emission,
          and the ``anomalous microwave emission''),
          other interstellar carbon species
          (e.g., C$_{60}$, carbon chains, and possibly graphene 
          and carbon nanotubes)\cite{Garcia2011,Berne2012,Li2019Graphene,Chen2019CNT}, 
          and the chemical complexity of the Universe 
          including the formation of H$_2$
          and their role as a viable sink of interstellar 
          deuterium\cite{Draine2006}?
\item How do PAHs evolve from the ISM 
          to prestellar nebulae, protoplanetary disks,
          %and small solar system bodies 
          %(comets, asteroids, meteorites, 
          %and interplanetary dust particles)?
          %Why PAHs are not seen in embedded protostars,
          %the early phases of star formation? 
          %why are PAHs generally smaller in 
          %meteorites, comets,  
          %interplanetary dust particles\cite{Derenne2010}
          %and T Tauri disks\cite{Seok2017} than that in the ISM? 
          comets, and meteorites? 
          Why PAHs are not seen in embedded protostars,
          the early phases of star formation? 
          why are PAHs generally smaller in 
          meteorites\cite{Derenne2010}
          and T Tauri disks\cite{Seok2017} than that in the ISM? 
\end{itemize}
}
\end{minipage}
}%
\end{figure*}

\section*{Concluding Remarks}
{\it Spitzer} has provided a wealth of spectral 
and imaging data on the characteristics and 
spatial distributions of PAHs in a wide variety 
of astrophysical regions and considerably 
expanded the field of PAH astrophysics, 
especially the extragalactic world 
up to $z$\,$\simgt$\,4 (see Box~1).
The PAH model has been successful 
in explaining the overall spectral profiles 
and the band patterns
observed in various regions in terms of 
a mixture of neutral and charged PAHs 
of different sizes\cite{Allamandola1999,DL01,LD2001}.
%(e.g., see Allamandola et al.\ 1999, 
%Draine \& Li 2001, 2007, Li \& Draine 2001,
%Bauschlicher et al.\ 2010, Cami et al.\ 2011, 
%Rosenberg et al.\ 2011, Boersma et al.\ 2013, 2014, 
%Andrews et al.\ 2015).
%
However, longstanding open questions remain,
with some of which arising from {\it Spitzer}
observations (see Box~2).

{\it Spitzer}/IRS observations of
harsh environments such as 
%compact H{\sc ii} regions\cite{Stock2013},
AGN-hosting Seyfert galaxies\cite{Diamond2010}
and galaxy mergers\cite{Haan2011}
%are somewhat challenging for the PAH model
%to reproduce the observed extreme 11.3/7.7
find extreme 11.3/7.7
band ratios which lie beyond the range of model
predictions, even for completely neutral PAHs\cite{DL01}.
Future experimental measurements,
quantum chemical computations
and theoretical modeling 
of large catacondensed PAHs 
with an open, irregular structure 
will be useful to attest the PAH model.
%
%{\bf 
Andrews et al.\cite{Andrews2015}
suggested that the interstellar PAH family
may be dominated by ``grandPAHs'', 
large and hence stable molecules that can 
survive the harsh conditions of the ISM.
However, it is difficult for ``grandPAHs'' 
to reconcile with these extreme 11.3/7.7
band ratios (see Fig.\,\ref{fig:pah4agn}).
%}
%

%
%...posed a severe challenge against
%
%Another challenge posed against the PAH model
%is that, so far there is no actual precise identification 
%of a single specific PAH molecule
Another challenge to the PAH model
is that, so far, no specific PAH molecule 
has been identified
in the interstellar or circumstellar space\cite{KZ2011}.
%(see Kwok \& Zhang 2011).
%Details of the PAH spectra 
%(precise band positions, bandwidths, 
%and relative band intensities)
%remain hard to mimic exactly 
%with the use of available PAH spectra 
%obtained by experimental measurements 
%or quantum chemical calculations.
%
%{\bf
Laboratory measurements have shown that 
individual small PAH molecules 
(with $\simlt$\,25 C atoms) have strong and narrow 
absorption features in the UV\cite{Salama1995}.  
Attempts to search for these features were made 
with the {\it Space Telescope Imaging Spectrograph} 
on board the {\it Hubble Space Telescope}\cite{Clayton2003}.
However, no such absorption features were seen. 
%Andrews et al.\cite{Andrews2015}
%suggested that the interstellar PAH family
%may be dominated by ``grandPAHs'', 
%large and hence stable molecules that can 
%survive the harsh conditions of the ISM.
%
We argue that it is natural to expect a large number 
of distinct PAH species to be present in the ISM, and
no single UV band may be strong enough to be identified.
The strong interstellar 2175$\Angstrom$ extinction feature 
is likely to be a blend of $\pi$--$\pi^{\ast}$ absorption bands
from the entire population of 
PAHs\cite{Joblin1992,LD2001,Cecchi2008,Steglich2011,Mulas2013}. 
Furthermore, internal conversion may lead to 
extreme broadening of the UV absorption bands 
in larger PAHs, which may account for absence of
recognizable UV absorption features 
other than the 2175$\Angstrom$ bump.
%}
%shortward of 
%$\simali$2000$\Angstrom$.
%
Therefore, the lack of identification of 
any specific PAH is not a fatal problem 
for the PAH model, at least at this time. 
As we develop a better knowledge 
of the gas-phase spectroscopy of larger PAHs, 
this may change. 
If the diffuse interstellar bands are 
electronic transitions of PAHs, 
they hold great promise for identifying 
specific PAH molecules, as the electronic transitions 
are more characteristic of a specific PAH molecule 
than the mid-IR C--H and C--C vibrational bands,
while the latter are mostly representative of 
functional groups and thus do not allow one 
to fingerprint individual PAH molecules. 

It remains unclear where and how interstellar PAHs 
are formed. Suggested sources for interstellar
PAHs include the formation in the ISM through 
gas-phase ion-molecule reactions, 
%\cite{Herbst1991} 
or through shattering of hydrogenated amorphous 
carbon dust by grain-grain collisions 
in interstellar shocks. 
%\cite{Scott1996}, 
Also suggested is the formation of PAHs
in carbon star outflows followed by  
subsequent injection into the ISM.
%\cite{Latter1991}. 
Nevertheless, the PAH emission bands are rarely 
seen in the mid-IR spectra of carbon stars. 
%{\bf
To our knowledge, only three carbon stars ---
TU Tau\cite{Buss1991,Speck1997,Boersma2006},
UV Aur\cite{Speck1997},
and HD\,100764\cite{Sloan2007} ---
are known to exhibit PAH emission.
%}
%Both TU Tau and UV Aur have a hot companion
%which emits UV photons\cite{Buss1991,Speck1997,Boersma2006}.
It has been suggested that PAHs are present in all C stars 
but they are simply not excited sufficiently 
to emit at mid-IR due to lack of UV photons,
noting that both TU Tau and UV Aur 
have a hot companion 
which emits UV photons\cite{Buss1991,Speck1997,Boersma2006}.
However, it has been shown that the excitation of PAHs 
do not need very energetic photons, 
instead, visible and near-IR photons are able to excite 
PAHs to temperatures high enough to emit the mid-IR 
bands\cite{LD2002,Mattioda2005},
in agreement with the detection of PAH emission
in regions lack of UV photons,
including vdB\,133, a reflection nebula
illuminated by a F5\,Iab star of effective temperature 
$\Teff$\,$\approx$\,6800$\K$\cite{Uchida1998},
HD\,100764, a C-rich red giant of 
$\Teff$\,$\approx$\,4850$\K$\cite{Sloan2007},
and HD\,233517, an O-rich red giant 
of $\Teff$\,$\approx$\,4475$\K$\cite{Jura2006}.
%

%{\bf
The distribution of PAHs in the SMC, 
derived from the {\it Spitzer}/IRAC 
photometric mapping data,
does not follow that of C-rich AGB stars,
instead, it correlates with molecular gas
as traced by CO (ref.\cite{Sandstrom2010}). 
This suggests that PAHs may be forming 
in molecular clouds.
In contrast, 
the distribution of PAHs in the LMC, 
also derived from the {\it Spitzer}/IRAC 
mapping data,
is enhanced both in molecular clouds
and in the stellar bar, which hosts 
the highest concentration of 
AGB stars\cite{Paradis2009}.
The association of PAH emission
with the stellar bar in the LMC 
was considered as evidence 
for C-rich AGB stars 
as a PAH source.
However, we argue that,
if PAHs are formed from 
the fragmentation of carbonaceous 
dust grains which were condensed in AGB stars 
and subsequently injected into the ISM,
one would also expect the distribution of PAHs 
to be enhanced in the stellar bar. 
%}
%
With the Faint Object Infrared Camera 
%for the SOFIA Telescope 
(FORCAST) 
onboard the Stratospheric Observatory 
for Infrared Astronomy (SOFIA), 
%The SOFIA/FORCAST 
the mid-IR imaging observations of 
NGC\,7027, a planetary nebula of age
$\simali$1000$\yr$, suggested rapid PAH formation 
along the outflow via grain-grain collisions 
in the post-shock environment 
of the dense photodissociation region 
and molecular envelope\cite{Lau2016}.
With the upcoming JWST, smaller spatial scales 
can be probed; spectral mapping in the PAH bands
of the outflows of C-rich AGB stars 
and planetary nebulae will be valuable 
for exploring the origin and evolution of PAHs.

The emission continuum
underlying the discrete PAH features
is pervasive\cite{Xie2018nano}
and its nature remains unclear\cite{Kwok2001}.
It is noteworthy that the reliability of 
PAH emission as an effective SFR indicator 
relies on 
%%robust measurements 
%%of the PAH features 
the appropriate subtraction
of the underlying continuum
for which no consensus has 
yet been reached\cite{Uchida2000,Rapacioli2005,Smith2007,Peeters2017,Xie2018PAH}.
The observed band ratios
(e.g., the extreme 11.3/7.7 ratios 
seen in hostile environments)
used to derive the PAH size, charge
and structure and the local physical
and chemical conditions also
depend on how the continuum
is defined and subtracted. 
The continuum emission was posited 
to be related to the anharmonicity of
highly vibrationally excited PAHs\cite{ATB1989},
some specific PAH species (e.g., tubular PAHs)
with a zero bandgap\cite{DL07},
or energetically processed PAHs\cite{Cruz-Diaz2019}.
%
%Anharmonic interactions may also lead 
%to a vibrational quasi continuum.
%
%Alternatively, the delayed electronic 
%fluorescence process pointed out by 
%Hansen \& Ferrari may result in 
%a near-IR continuum.
%
Kwok \& Zhang\cite{KZ2011}
proposed that organic nanoparticles 
with a mixed aromatic-aliphatic structure 
could be responsible for 
both the PAH emission bands 
and the underlying continuum. 
However, the 3.3$\mum$ emission feature 
and the underlying continuum emission 
at $\simali$2$\mum$ in NGC\,2023, 
a reflection nebula, are spatially separated, 
suggesting the 3.3$\mum$ feature and
the underlying continuum do not share
the same carrier\cite{An2003}.
Future spatially resolved observations 
with JWST and SPICA
that can map the two components 
will provide insight into their physical separations 
and chemical carriers.
%

%Admittedly, the spectroscopic capabilities 
%of {\it Spitzer} were somewhat limited, both
%in spectral resolution and wavelength coverage.
Compared with {\it Spitzer}, JWST will have more 
than an order of magnitude increase in sensitivity 
and spatial resolution as well as a broader wavelength
coverage in the near-IR.
While the 3.3$\mum$ C--H stretch 
and the accompanying satellite features
at 3.4--3.6$\mum$ are beyond 
the wavelength range of {\it Spitzer}/IRS, 
it is expected that JWST, with its 
{\it Near InfraRed Spectrograph} (NIRSpec) 
operating at 0.6--5$\mum$,
will be able to scrutinize these bands
so as to study the chemical structures
(e.g., methylation, superhydrogenation, 
and anharmonicity) of the smallest PAHs 
and their environmental dependence. 
Moreover, JWST/NIRSpec will be ideal
for searching for the overtone band 
at $\simali$1.6--1.8$\mum$ 
of the 3.3$\mum$ feature of
small PAHs, providing valuable 
insights into the PAH fluorescence 
process\cite{Geballe1994,Chen2019Anharm}. 
Furthermore, the C--D stretch at $\simali$4.4$\mum$
of deuterated PAHs will be of particular interest.
PAHs could be a major reservoir of deuterium in the ISM. 
PAHs of intermediate size (with $\simlt$\,100 C atoms)
are expected to become deuterium enriched in the ISM 
through the selective loss of hydrogen 
during photodissociation events\cite{Draine2006}.
While ISO/SWS and AKARI observations
have reported the tentative detection of 
the C--D bands at 4.4 and 4.65$\mum$ 
in the Orion Bar and M17 PDRs\cite{Peeters2004dPAH,Onaka2014} 
and H{\sc ii} regions\cite{Doney2016},
the unique high sensitivity of JWST/NIRSpec
will place the detection of deuterated PAHs 
on firm ground and enable far more detailed 
band analysis than previously possible.

The high sensitivity and high spatial resolution 
capabilities of JWST and SPICA
will open up an IR window unexplored by 
{\it Spitzer} and unmatched by {\it ISO} observations. 
%due to the sensitivity difference.
%and thus will place the detection of PAHs 
%in faint sources and distant objects 
%on firm ground and enable far more detailed band
%analysis than previously possible.
With JWST and SPICA aided by laboratory studies, 
quantum chemical computations, 
and theoretical modeling, 
the future of PAH astrophysics 
will be even more promising!

%\bibliography{review}
%\bibliography{review}

\section*{Acknowledgements}
I dedicate this article to the 60th anniversary 
of the Department of Astronomy 
of Beijing Normal University, 
the 2nd astronomy program 
in the modern history of China.
I thank B.T.~Draine, L.C.~Ho, M.~Karouzos, 
X.J.~Yang and the three anonymous referees
for very useful comments and suggestions.
I thank L.~Armus, P.~Beir\~{a}o, 
J.G.~Ingalls, H.~Kaneda, D.~Lutz,
K.~Mattila, D.A.~Riechers, B.~Siana, O.~Vega,
M.~Yamagishi, and L.~Yan for providing 
the PAH spectra shown in 
Figs.\,\ref{fig:pahspec}--\ref{fig:m82pah}.
I thank X.Z.~Chen, B.~Yang and W.B.~Zuo 
for their help during the manuscript preparation.
This work is supported in part by NASA grants 
80NSSC19K0572 and 80NSSC19K0701.

\section*{Competing interests}
The author declares no competing financial interests.

\section*{Additional information}
\noindent{Correspondence should be addressed to A.L.}

\end{document}